%% file: SCD_low_dissipation_cavity_optomechanics_arXiv.tex
\documentclass[aps,pra,superscriptaddress,twocolumn]{revtex4-1}
\pdfoutput=1
\usepackage[version=3]{mhchem} 
\usepackage[dvips]{epsfig}
\usepackage{graphicx}  
\usepackage{dcolumn}   
\usepackage{bm}        
\usepackage{amssymb}   
\usepackage{color}
\usepackage{float}
\usepackage{hyperref}
\hypersetup{
     colorlinks   = true,
     citecolor    = blue,
     linkcolor    = red,
     urlcolor     = black
     }
\usepackage{etoolbox}
\patchcmd{\section}
  {\centering}
  {\raggedright}
  {}
  {}
\patchcmd{\subsection}
  {\centering}
  {\raggedright}
  {}
  {}

\begin{document}

\author{Matthew Mitchell}
\affiliation{Department of Physics and Astronomy and Institute for Quantum Science and Technology, University of Calgary, Calgary, AB, T2N 1N4, Canada}
\affiliation{National Institute for Nanotechnology, Edmonton, AB, T6G 2M9, Canada}

\author{Behzad Khanaliloo}
\affiliation{Department of Physics and Astronomy and Institute for Quantum Science and Technology, University of Calgary, Calgary, AB, T2N 1N4, Canada}
\affiliation{National Institute for Nanotechnology, Edmonton, AB, T6G 2M9, Canada}

\author{David P. Lake}
\affiliation{Department of Physics and Astronomy and Institute for Quantum Science and Technology, University of Calgary, Calgary, AB, T2N 1N4, Canada}
\affiliation{National Institute for Nanotechnology, Edmonton, AB, T6G 2M9, Canada}

\author{Tamiko Masuda}
\affiliation{Department of Physics and Astronomy and Institute for Quantum Science and Technology, University of Calgary, Calgary, AB, T2N 1N4, Canada}

\author{J.P. Hadden}
\affiliation{Department of Physics and Astronomy and Institute for Quantum Science and Technology, University of Calgary, Calgary, AB, T2N 1N4, Canada}

\author{Paul E. Barclay}
\email{pbarclay@ucalgary.ca}
\affiliation{Department of Physics and Astronomy and Institute for Quantum Science and Technology, University of Calgary, Calgary, AB, T2N 1N4, Canada}
\affiliation{National Institute for Nanotechnology, Edmonton, AB, T6G 2M9, Canada}

\title{Single-crystal diamond low-dissipation cavity optomechanics}

\begin{abstract}
 Single-crystal diamond cavity optomechanical devices are a promising example of a hybrid quantum system: by  coupling  mechanical resonances to both light and electron spins, they can enable new ways for photons to control solid state qubits. However, realizing cavity optomechanical devices  from high quality  diamond chips has been an outstanding challenge. Here we demonstrate single-crystal diamond cavity optomechanical devices that can enable photon-phonon-spin coupling. Cavity optomechanical coupling to $2\,\text{GHz}$ frequency ($f_\text{m}$) mechanical resonances is observed.  In room temperature ambient conditions, these resonances have a record combination of low dissipation (mechanical quality factor, $Q_\text{m} > 9000$) and high frequency, with $Q_\text{m}\cdot f_\text{m} \sim 1.9\times10^{13}$ sufficient for room temperature single phonon coherence. The system exhibits high optical quality factor ($Q_\text{o} > 10^4$)  resonances at infrared and visible wavelengths, is nearly sideband resolved, and exhibits optomechanical cooperativity $C\sim 3$. The devices' potential for optomechanical control of diamond electron spins is demonstrated through radiation pressure excitation of mechanical self-oscillations whose 31 pm amplitude is predicted to provide 0.6 MHz coupling rates to diamond nitrogen vacancy center ground state transitions (6 Hz / phonon), and $\sim10^5$ stronger coupling rates to excited state transitions.
\end{abstract}

\maketitle

\section{Introduction}
\noindent Diamond cavity optomechanical devices are an attractive platform for controlling interactions between light, vibrations, and electrons that underly future hybrid quantum technologies  \cite{ref:treutlein2014hms}. Their potential arises from diamond's exceptional mechanical and optical properties \cite{ref:aharonovich2011dp} combined with its ability to host color centers such as the nitrogen-vacancy (NV) whose electron spins are excellent qubits that can be manipulated by local mechanical strain fields \cite{ref:macquarrie2013msc, ref:teissier2014scn, ref:ovartchaiyapong2014dsc, ref:barfuss2015smd,ref:meesala2016esc}. Recently, piezoelectric actuation of  bulk \cite{ref:macquarrie2013msc, ref:macquarrie2015ccn} and nanomechanical \cite{ref:arcizet2011snv,ref:teissier2014scn,ref:ovartchaiyapong2014dsc,ref:barfuss2015smd, ref:meesala2016esc,ref:golter2016oqc,ref:lee2016scm} diamond resonators has been used to demonstrate phononic spin control. Cavity optomechanics  \cite{ref:aspelmeyer2014co} harnesses optical forces in place of piezoelectric actuation, allowing coherent phonon state manipulation  \cite{ref:weis2010oit, ref:safavi2011eit, ref:liu2013eit} of GHz frequency mechanical resonators with quantum limited sensitivity \cite{ref:chan2011lcn}. These phonons can be made resonant with NV center electron spin transitions that are central to proposals for spin--spin entanglement  \cite{ref:kepesidis2013pcl}, spin--phonon state transfer  \cite{ref:rabl2010qst, ref:stannigel2010otl, ref:schuetz2015uqt}, spin mediated mechanical normal mode cooling  \cite{ref:wilson2004lcn, ref:kepesidis2013pcl,ref:macquarrie2016cmr}, and photon-phonon-spin coupling \cite{ref:ramos2013nqo}. Additionally, the relatively small thermal occupancy and mechanical dissipation of GHz diamond devices, combined with diamond's ability to support strong optical fields due its large electronic bandgap, make them an ideal system for the cavity optomechanical backaction cooling and study of  mechanical resonators in their quantum ground state \cite{ref:chan2011lcn}.

\begin{figure*}[ht]
\epsfig{figure=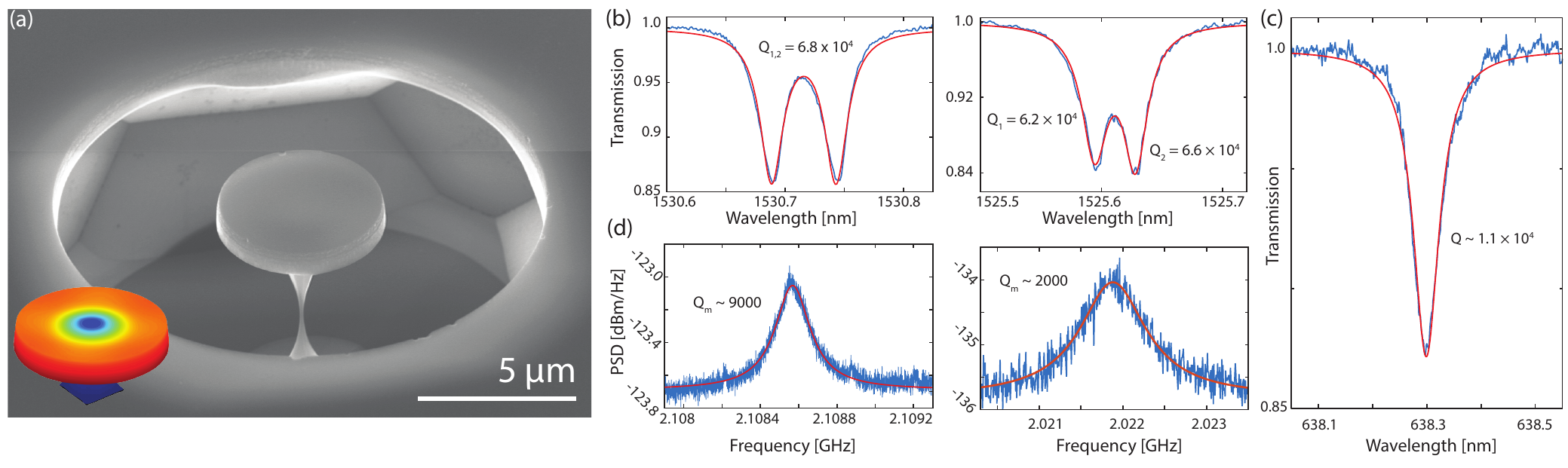, width=1\linewidth}
 \caption{Characterization of diamond microdisk optical and mechanical modes at low optical input power. (a) SEM image of a 5 $\mu$m diameter microdisk, with minimum pedestal width of 100 nm. Inset: Simulated displacement distribution of the RBM mechanical resonance of this device.  (b) Highest $Q_\text{o}$ TM--like optical modes of a 5 $\mu$m (left) and 5.5 $\mu$m (right) diameter microdisk, with intrinsic quality factors for each doublet resonances as labeled. (c) High--$Q_\text{o}$ visible mode, with intrinsic quality factor as shown for a $6.2\,\mu\text{m}$ diameter microdisk. (d) $S_P(f)$ produced by the thermal motion of the RBM of the microdisks in (b), showing that the larger diameter, larger pedestal waist microdisk has a lower $Q_\text{m}$.}
\label{fig:SEM}
\end{figure*}

Development of cavity optomechanical devices from single-crystal diamond has been limited due to challenges associated with fabricating mechanically isolated structures from bulk diamond chips. While initial development of diamond optomechanical devices used nanocrystalline diamond \cite{ref:rath2013dio}, single crystal diamond material promises lower mechanical dissipation  \cite{ref:tao2013scd} and the ability to host highly coherent NV centers  \cite{ref:balasubramanian2009usc}. Here we report demonstration of a single-crystal diamond cavity optomechanical system for the first time. This system is based on a microdisk device geometry that has been used in a wide range of cavity optomechanics experiments implemented in more conventional semiconductor and dielectric materials \cite{ref:weis2010oit, ref:liu2013eit, ref:gil2015hfn, ref:lu2015hfh}. Microdisks are desirable owing to their simple geometry, strong optomechanical coupling between high frequency mechanical resonances and low loss optical modes, and intrinsic ability to simultaneously support optical modes over the entire transparency window of the device material \cite{ref:liu2013eit}.

The microdisk system studied here, an example of which is shown in  Fig.\ \ref{fig:SEM}(a), supports optical modes at visible and telecommunication wavelengths ($\omega/2\pi \sim 200 - 470$ THz) that interact via radiation pressure with GHz frequency mesoscopic mechanical resonances of the structure. We find that these resonances have a record combination of high $\omega_\text{m}$ and low mechanical dissipation ($\gamma_\text{m} = \omega_\text{m}/Q_\text{m} \sim 2\pi\times 0.2\,\text{MHz} $) compared to other mechanical resonators operating in ambient temperature and pressure, and that their $Q_\text{m}\cdot f_\text{m} = 1.9 \times 10^{13}$ Hz product is sufficiently high to satisfy the minimum criteria for single phonon coherent behaviour \cite{ref:aspelmeyer2014co}. The microdisk optical modes have low dissipation ($\gamma_\text{o} = \omega_\text{o}/Q_\text{o}\sim 2\pi\times 3\,\text{GHz}$), and owing to the negligible nonlinear absorption in diamond at telecom optical frequencies, they can support intracavity photon number $N > 10^6$ without degrading $Q_\text{o}$. In combination, this allows realization of optomechanical cooperativity, $C = N g_0^2/\gamma_\text{o}\gamma_\text{m} \sim 3$, large enough ($>1$) for coherent photon-phonon coupling \cite{ref:weis2010oit, ref:safavi2011eit}, where $g_0 \sim 2\pi\times 26\,\text{kHz}$ is the single photon optomechanical coupling rate of the device and describes the expected shift in cavity optical frequency due to the mechanical zero point motion of the microdisk. These devices operate on the border of the sideband resolved regime ($\gamma_\text{o} \sim \omega_\text{m}$), enabling radiation pressure backaction excitation of mechanical self-oscillations with $\sim 31\,\text{pm}$ amplitude. The accompanying stress fields are strong  enough to drive diamond color center spin transitions with a single phonon-spin coupling rate that is predicted to exceed that of previously studied MHz frequency nanomechanical structures  \cite{ref:barfuss2015smd, ref:ovartchaiyapong2014dsc, ref:meesala2016esc}, despite having orders of magnitude higher $\omega_\text{m}$ and smaller phonon amplitude, owing to the localized nature of the microdisk mechanical resonances. In addition, the ability of the microdisks to support optical modes at visible wavelengths is compatible with resonant coupling to NV center optical transitions \cite{ref:faraon2011rez}, as well as operation in fluid environments of interest for sensing applications \cite{ref:gil2015hfn}.

\section{Fabrication of single-crystal diamond microdisks}
\noindent There has been significant recent progress in fabrication of mechanically isolated single-crystal diamond devices, including demonstrations of suspended high-$Q_\text{m}$ nanomechanical resonators  \cite{ ref:teissier2014scn,ref:ovartchaiyapong2012hqf,ref:burek2012fsm,ref:tao2013scd, ref:khanaliloo2015dnw} and  high-$Q_\text{o}$ micro- and nanocavities  \cite{ref:burek2014hqf, ref:khanaliloo2015hqv}. These structures have been created using  diamond membrane thinning  \cite{ref:maletinsky2012ars,ref:ovartchaiyapong2012hqf,ref:tao2013scd}, plasma angled-etching  \cite{ref:burek2014hqf}, and plasma undercutting \cite{ref:khanaliloo2015dnw, ref:khanaliloo2015hqv} fabrication techniques, with the latter two approaches allowing patterning of devices from bulk diamond chips. Here we use plasma undercutting to fabricate single crystal diamond cavity optomechanical devices \cite{ref:khanaliloo2015dnw, ref:khanaliloo2015hqv}. These devices were fabricated from an optical grade, chemical vapor deposition (CVD) grown $\langle 100 \rangle$-oriented SCD substrate supplied by Element Six. The polished substrates were first cleaned in boiling piranha, and coated with $\sim$ 400 nm of plasma-enhanced chemical vapor deposition (PECVD) Si$_3$N$_4$ as a hard mask. To avoid charging effects during electron beam lithography (EBL), $\sim$ 5 nm of Ti was deposited on the Si$_3$N$_4$ layer, before coating the sample with the ZEP 520A EBL resist. The developed pattern was transferred to the hard mask via inductively coupled reactive ion etching (ICPRIE) with C$_4$F$_8$/SF$_6$ chemistry. The remaining EBL resist was removed with a 6 minute deep--UV exposure ($5\ \text{mW}/\text{cm}^2$ at 254 nm) followed by a 2 minute soak in Remover PG, while the remaining Ti was removed by the subsequent etch steps. The anisotropic ICPRIE diamond etch was performed using O$_2$, followed by deposition of $\sim$ 250 nm of conformal PECVD Si$_3$N$_4$ as a sidewall protection layer. The bottom of the etch windows were then cleared of Si$_3$N$_4$ using a short ICPRIE C$_4$F$_8$/SF$_6$ etch. This was followed by a zero RF power O$_2$ RIE diamond undercut etch to partially release the devices. Lastly, the Si$_3$N$_4$ layer was removed using a wet-etch in 49$\%$ HF, and the devices were cleaned again in boiling piranha. The devices studied here have diameters of $5.0\,\mu\text{m}$ to $6.0\,\mu\text{m}$ and average thickness $\sim 940\,\text{nm}$. As evident from the image in Fig.\ \ref{fig:SEM}(a),  devices are fabricated with a process optimized to minimize the waist of the pedestal supporting the microdisk, reducing it to $< 100\,\text{nm}$, where the waist is defined as the smallest point of the pedestal. The microdisk thickness, which will be reduced in future work to enhance confinement, is determined by the interplay between the inward and upward etch rates of the quasi-isotropic undercut, together the initial anisotropic etch depth. The undercut time was chosen to optimize the pedestal waists of the $\sim 5 \ \mu\text{m}$ diameter disks studied here. A longer undercut would allow the study of larger diameter microdisks (6 $\mu\text{m}$ to 8 $\mu\text{m}$) present on the chip, which would in turn possess a smaller thickness than the structures studied here.

\noindent\section{Device characterization}
\subsection{OPTICAL CHARACTERIZATION}

\noindent The devices were characterized by monitoring the transmission of a dimpled optical fiber taper \cite{ref:michael2007oft,ref:mitchell2014cog} evanescently coupled to the microdisk and input with light from tunable diode lasers (New Focus Velocity) with  wavelengths near 1530 nm or 637 nm. For the 1530 nm measurements, the output of the fiber taper was monitored by both low- and high-bandwidth photoreceivers (Newport 1621 and 1554-B, respectively), and a calibrated optical power meter (Newport 2936-R). Figure \ref{fig:SEM}(b) shows typical $\overline{T}(\lambda_\text{s})$ when the fiber taper is evanescently coupled to devices with diameters of $5.0\,\mu\text{m}$ and $5.5\,\mu\text{m}$ and the wavelength $\lambda_\text{s}$  of the 1530 nm tunable laser  is scanned across microdisk modes at $\lambda_\text{o}$.  Here $\overline{T}$ is the average transmission measured by the low-bandwidth photodetector over a timescale long compared to $1/f_\text{m}$. These measurements reveal resonant coupling to modes with loaded $Q_\text{o} \sim 5.8\times10^4 - 6.0\times 10^4$ (intrinsic $Q_\text{o}^{(i)} = 6.1\times10^4 - 6.8\times10^4$), and a degree of doublet structure that depends on the internal backscattering of a given device. Maximizing $Q_\text{o}$, and thereby minimizing $\gamma_\text{o}$, is important for achieving the aforementioned regime allowing coherent photon--phonon coupling ($C > 1$).

The ability of these devices to support modes over a wide wavelength range is demonstrated in Fig.\,\ref{fig:SEM}(c), where  the 637 nm tunable laser was used to probe a mode with high-$Q_\text{o} > 1\times10^4$. This is promising for applications involving NV center optical transitions in this wavelength range \cite{ref:golter2016oqc}. These devices have a predicted radiation loss limited $Q_\text{o} > 10^7$ at both 1550 nm and 637 nm wavelengths; $\gamma_\text{o}$ is currently limited by surface roughness and linear absorption. In previous work the microdisk pedestal size was observed to limit $Q_\text{o}$ for insufficient relative undercut. In the devices studied here scattering due to the non--cylindrical pedestal shape is predicted to dominate the contribution to $\gamma_\text{o}$ from the pedestal \cite{ref:khanaliloo2015hqv}. The lower $Q_\text{o}$ observed at 637 nm can be attributed in part to sub--optimal fiber taper positioning \cite{ref:borselli2004rsm} and an increased sensitivity to surface scattering \cite{ref:burek2014hqf}.

\subsection{CAVITY OPTOMECHANICAL COUPLING}

\noindent To probe optomechanical coupling within the microdisks, time ($t$) dependent transmission fluctuations $\delta T(t,\lambda_\text{s})$ of $1550~\text{nm}$ light  were monitored using a real time spectrum analyzer (Tektronix RSA5106A). Excitations of the microdisk mechanical resonances modulate $\lambda_\text{o}$, resulting in a dispersive optomechanical transduction of mechanical motion to an optical signal ${P_o}\delta T(t;\lambda_\text{s})$ that can then be observed in the measured electronic power spectrum $S_P(f)$. Here ${P_o}$ is the average power transmitted to the photoreceiver. Figure \ref{fig:SEM}(d) shows typical spectra when $\lambda_\text{s}$ is tuned near the point of maximum transduction of the modes in Fig.\ \ref{fig:SEM}(b). Resonances near $f_\text{m} \sim 2.0 - 2.1\,\text{GHz}$ are observed, corresponding to optomechanical transduction of the thermomechanical motion of the fundamental radial breathing mode (RBM) of the microdisks.  The predicted displacement of the RBM calculated using finite element simulations (COMSOL) is shown in the inset to Fig.\ \ref{fig:SEM}(a). The simulated $f_\text{m}$ of the RBM for varying microdisk diameter was found to be within 10\% of observed values.  These measurements were conducted at low input power $P_i \sim 50 \ \mu$W to avoid optomechanical backaction effects discussed below. Here an erbium doped fiber amplifier (EDFA: Pritel LNHPFA-30) was used on the output side of the fiber taper to boost the optical signal prior to photodetection to a level just below the detector saturation power ($P_o \approx 0.7$ mW).

The microdisk pedestal can significantly affect the RBM properties, and minimizing its waist size is important in order to maximize $Q_\text{m}$ and reach $C > 1$. In previously studied diamond microdisks with $\mu$m  pedestal waists  \cite{ref:khanaliloo2015hqv}, transduction of mechanical modes was not observed. As shown in Fig.\ \ref{fig:SEM}(a), the devices used here for cavity optomechanics have significantly smaller waists, e.g., the $5.0\,\mu\text{m}$ diameter microdisks have waist $< 100\,\text{nm}$.  Figure\ \ref{fig:SEM}(d) shows that when microdisk diameter, and as a result, pedestal waist are increased to $5.5\,\mu\text{m}$ and $400\,\text{nm}$, respectively, $Q_\text{m}$ is found to decrease from $\sim 9000$ to $\sim 2000$. Mechanical resonances are not observed in devices with pedestal diameter $> 500$ nm. This indicates that $Q_\text{m}$ for these devices is limited by clamping loss  \cite{ref:nguyen2013uqf, ref:lu2015hfh}. The hourglass shape of the pedestals obtained for the $\langle 100\rangle$ optical grade diamond samples used here limits the minimum size of the pedestal where it connects to the microdisk, and may result in increased dissipation. Given the crystal plane selective nature of the diamond undercut  \cite{ref:khanaliloo2015hqv}, fabricating microdisks from samples with alternate crystal orientation such as $\langle 111\rangle$ may alleviate this limitation. Additionally, operation in vacuum, where viscous air damping can be avoided  \cite{ref:riviere:2011cos, ref:tang2012hqs}, and at low temperature  \cite{ref:khanaliloo2015dnw, ref:tao2013scd} would allow a decrease in dissipation, as the total $Q_\text{m}$ is given by $Q_\text{m} = (\sum_j1/Q_m^j)^{-1}$, where the $Q_m^j$ represent the quality factor due to each damping mechanism. Despite present limitations, the demonstrated devices have $Q_\text{m}\cdot f_\text{m} = 1.9\times 10^{13}$ Hz, that is larger than all previously studied cavity optomechanical systems operating in ambient conditions  \cite{ref:nguyen2013uqf, ref:lu2015hfh, ref:eichenfield2009oc}. A comparison of some of the highest $Q_\text{m}\cdot f_\text{m}$ products for optomechanical systems observed in ambient, cryogenic, and low pressure environments is shown in Supplement 1, Section 3. This figure of merit is critical for cavity optomechanical mass spectroscopy \cite{ref:ekinci2004uli, ref:yu2015cot, ref:gil2015hfn} and low phase noise oscillators \cite{ref:nguyen2007mtt}. Within the context of quantum optomechanics, this product satisfies a key minimum requirement for room temperature studies of single phonon coherence by over an order of magnitude:  $\omega_\text{m}/\gamma_\text{m} \gg n_\text{th}$, where $n_\text{th}$ is the room temperature phonon population of the RBM, ensuring that thermal decoherence is slow compared to a mechanical oscillation \cite{ref:aspelmeyer2014co}. By satisfying this condition, cooling to the quantum ground state from room temperature should also be possible \cite{ref:norte2016mrq}.

\begin{figure}[h]
\epsfig{figure=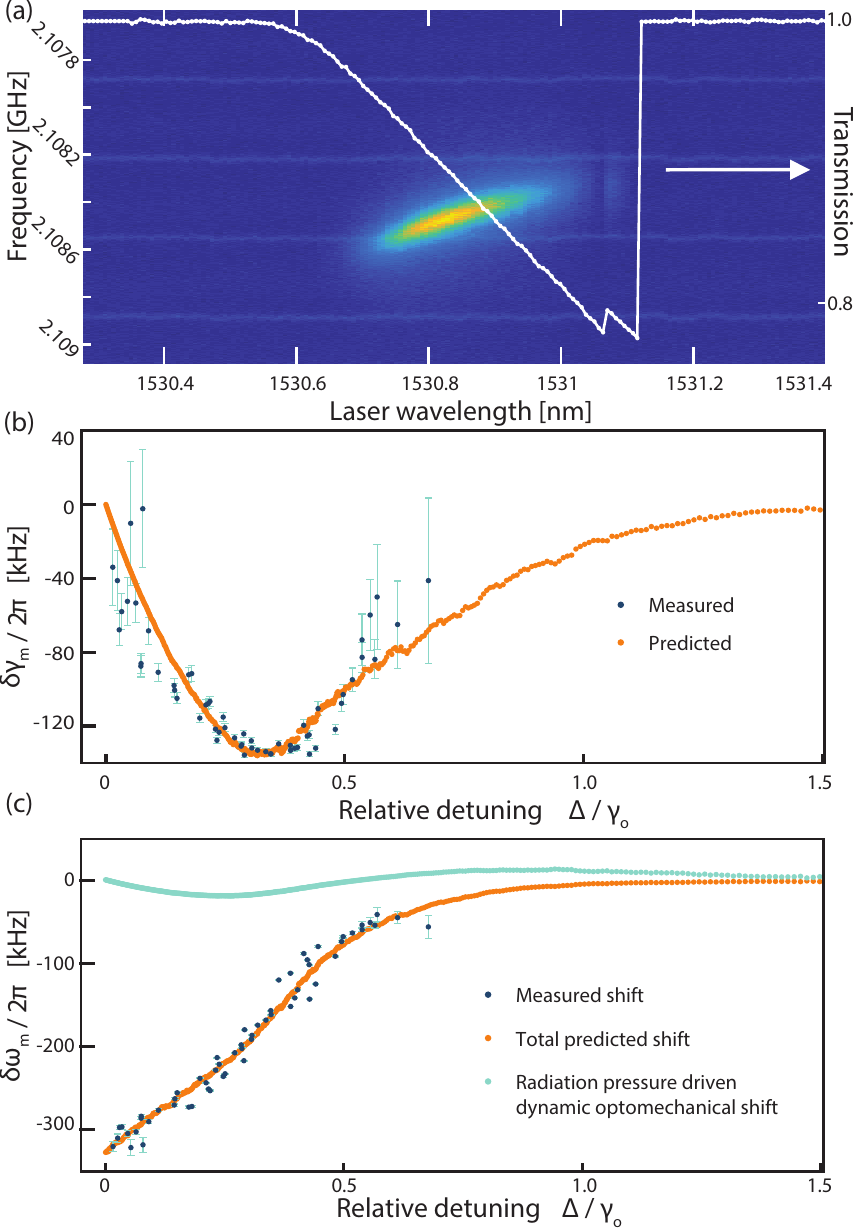, width=1\linewidth}
 \caption{Optomechanical backaction measuerments. (a) $S_P(f,\lambda_\text{s})$ and corresponding fiber transmission $T(\lambda)$ for $P_i$ corresponding to maximum $N \sim 6.5 \times 10^5$ and $P_d \sim 1.5$ mW. The regularly spaced horizontal features are electronic noise from the apparatus. (b) Observed and predicted optomechanical linewidth narrowing of the RBM. The predicted $\delta\gamma_\text{m}$ depends on measured $N$ and $\Delta$ for each point, as well as fitting parameter $g_0/2\pi = 26\pm2\,\text{kHz}$. Error bars indicate 95\% confidence interval for $\gamma_\text{m}$ extracted from $S_P(f)$ at each data point. (c) Observed and predicted $\delta\omega_m$. Both the predicted shift due to optomechanical backaction for $g_0$ found from the fits in (b), and the predicted shift including an additional static thermal softening determined by a free fitting parameter, are shown.}
\label{fig:optomechanics}
\end{figure}

To investigate the response of the cavity optomechanical transduction, $S_P$ was monitored while $\lambda_\text{s}$ was scanned across $\lambda_\text{o}$. Figure \ref{fig:optomechanics}(a) shows the resulting measurement of $S_P(f, \lambda_\text{s})$ for the microdisk in Fig.\ \ref{fig:SEM}(a), clearly illustrating that optomechanical transduction is only observable when $\lambda_\text{s}$ is tuned in the vicinity of $\lambda_o$.  In this measurement the EDFA was connected to the input side of the fiber taper,  resulting in maximum $N \sim 6.5 \times 10^5$ and $P_d \sim 1.5$ mW, where $P_d$ is the optical power dropped into the microdisk mode. From the thermo--optic coefficient of diamond we estimate that the shift of 400 pm from the cold cavity $\lambda_\text{o}$, as seen in Figure \ref{fig:optomechanics}(a), corresponds to a change in device temperature $\Delta\text{T} \sim 50$ K. COMSOL simulations that take into account the reduced thermal conductivity coefficient of the $\sim 100$ nm diameter pedestal compared to bulk \cite{ref:li2012tcd} confirm that an absorbed power of $\sim 10\%$ of $P_d$ reproduces the temperature shift observed here (see Supplement 1, Section 1). Although this temperature increase is not desirable for quantum optomechanics applications, it is considerably smaller than the expected increase for a similar Si device. Assuming a similar absorption rate and identical device geometry, a silicon device with a silicon or silicon dioxide pedestal would result in $\Delta\text{T} \sim 200$ K or 450 K, respectively, where a modified thermal conductivity also applies for the silicon in the pedestal \cite{ref:li2003tci}. It is expected that the rate of linear absorption observed here, which corresponds to $Q^{\text{abs}}_\text{o} = 6.2 \times10^5$, can be reduced through improvements to processing, as diamond devices with $Q_\text{o} > 10^6$ have been reported elsewhere \cite{ref:hausmann2014dnp}.

\subsection{CAVITY OPTOMECHANICAL BACKACTION}

\noindent The influence of cavity optomechanical backaction  \cite{ref:aspelmeyer2014co} on the dynamics of the mechanical resonator is analyzed in Figs.\ \ref{fig:optomechanics}(b,c).  Changes  $\delta \gamma_\text{m}$ and $\delta \omega_\text{m}$ to $\gamma_\text{m}$ and $\omega_m$, respectively, were measured as a function of source--cavity detuning $\Delta = \omega_\text{s} - \omega_\text{o}$. Their strong dependence on $\Delta$ clearly indicates that the mechanical dynamics are affected by the intracavity field. We were prevented from measuring significant $\delta \gamma_\text{m}$ and $\delta \omega_\text{m}$ for red-detuned wavelengths ($\Delta < 0$) due to the thermal bistability present in our system for large $P_d$, as shown in Fig.\ \ref{fig:optomechanics}(a).  As such, this study concentrated on blue-detuned wavelengths ($\Delta > 0$), however implementation of cavity stabilization techniques \cite{ref:liu2013eit} may allow this limitation to be overcome in the future, enabling more effective investigations of cavity sideband cooling \cite{ref:chan2011lcn,ref:aspelmeyer2014co}, and optomechanically induced transparency \cite{ref:weis2010oit, ref:safavi2011eit, ref:liu2013eit}.  Determining $\Delta$ for each data point in this analysis required accounting for the dependence of $\omega_\text{o}$ on $N$ due to the thermo-optic effect. For a given operating $\omega_\text{s}$, $\omega_\text{o}$ was predicted from $N$ (see Supplement 1, Section 1), where $N(\omega_\text{s})$ was determined from $\overline{T}(\omega_\text{s})$, $P_i$,  $Q_\text{o}^{(i)}$ and measurements of loss through the fiber taper and other elements of the apparatus.

To quantitatively investigate the role of radiation pressure on the mechanical resonance dynamics, the observed $\delta\gamma_\text{m}(\Delta)$ was fit to the expected cavity optomechanical damping rate  \cite{ref:aspelmeyer2014co}, with the single photon optomechanical coupling rate $g_0$  as the only free parameter. Using this method, the fit shown in Fig.\ \ref{fig:optomechanics}(b) was obtained for $g_0/2\pi \sim 26\,\text{kHz}$, with an associated 95\% confidence interval of $\pm2$ kHz. Errors bars for each $\delta\gamma_\text{m}$ data point in Fig.\ \ref{fig:optomechanics}(b) represent the 95\% confidence interval of fits used to extract $\gamma_\text{m}$ from $S_P$.    The large uncertainty as well as the discrepancy between the measured and predicted values  when $\Delta \sim 0$ or $\gg \gamma_\text{o}$ are due to the low signal to noise of $S_P$  in these regions. This low signal to noise of $S_P$ also prohibited measuring the optomechanical response for $\Delta \gg \gamma_\text{o}$. The fit value for $g_0$ has good agreement with $g_0$ predicted from COMSOL calculations that include both moving boundary (MB) and photoelastic (PE) contributions \cite{ref:chan2012ooc}. The predicted $g_0$ is dependent on the spatial overlap of the optical field and mechanical displacement profile, which varies for each optical mode. We find that the fit value of $g_0$ most closely agrees with the predicted coupling rate to the second order radial TM--like mode, with $g_{0_{PE}}/2\pi = 18\,\text{to}\,24\,\text{kHz}$ and $g_{0_{MB}}/2\pi = 16\,\text{kHz}$. In comparison, $g_{0_{MB}}/2\pi=  17\,\text{kHz}$ ($19\,\text{kHz}$) and $g_{0_{PE}}/2\pi = 29$ to $36\,\text{kHz}$ ($-24$ to $-26\,\text{kHz}$)  for the fundamental TM (TE) mode of the microdisk. This is consistent with measurements of the mode polarization in the fiber taper that indicated that the microdisk mode studied here is TM polarized. Note that the stated uncertainty in the predicted $g_{0_{PE}}$ is due to variations in reported PE coefficients of single-crystal diamond \cite{ref:hounsome2006pcd}.

Figure \ref{fig:optomechanics}(c) shows a similar analysis of $\delta\omega_\text{m}(\Delta)$, indicating that $\omega_\text{m}$ is softened by over $300\,\text{kHz}$ by the intracavity field. This shift is due to both optomechanical dynamical backaction and static thermal effects. For the operating regime and devices used here, dynamic thermal effects are expected to be below 5\% of the optomechanical radiation pressure dynamical backaction effects, and can be neglected \cite{ref:eichenfield2009oc}. However, static thermal effects are significant. Heating of the microdisk for large $P_d$ results in both thermal expansion and a change in Young's modulus, resulting in a shift to $\omega_\text{m}$ \cite{ref:nguyen2015iod,ref:clark2005hqu,ref:yang2014tdo}. This effect is linear in $P_d$, assuming that $Q_\text{o}$ is independent of power, i.e., nonlinear absorption is small. To compare the measured $\delta\omega_\text{m}(\Delta)$ with theory, we used a model that includes radiation pressure induced optomechanical dynamic backaction \cite{ref:aspelmeyer2014co} and a  static heating term linearly proportional to $P_d$:

\begin{align}\label{eq:wm}
\delta\omega_{\text{m}}(\Delta)= & g_{0}^{2}N\left(\frac{\Delta-\omega_{\text{m}}}{\gamma_\text{o}^{2}/4+(\Delta-\omega_{\text{m}})^{2}}+\frac{\Delta+\omega_{\text{m}}}{\gamma_\text{o}^{2}/4+(\Delta+\omega_{\text{m}})^{2}}\right)\nonumber\\
 + & \alpha P_d.
 \end{align}

The resulting fit of \eqref{eq:wm} to the measured $\delta\omega_\text{m}$ is shown in Fig.\ \ref{fig:optomechanics}(c) to have close agreement. Notably, this model reproduces the kink in $\delta\omega_\text{m}(\Delta)$ where the amplitude of the optomechanical contribution reaches a maximum and changes sign. This fit was obtained with $g_0$ fixed to the value extracted from the analysis of $\delta\gamma_\text{m}$ in Fig.\ \ref{fig:optomechanics}(b), and with $\alpha$ as a fitting parameter.

\begin{figure}[h]
\epsfig{figure=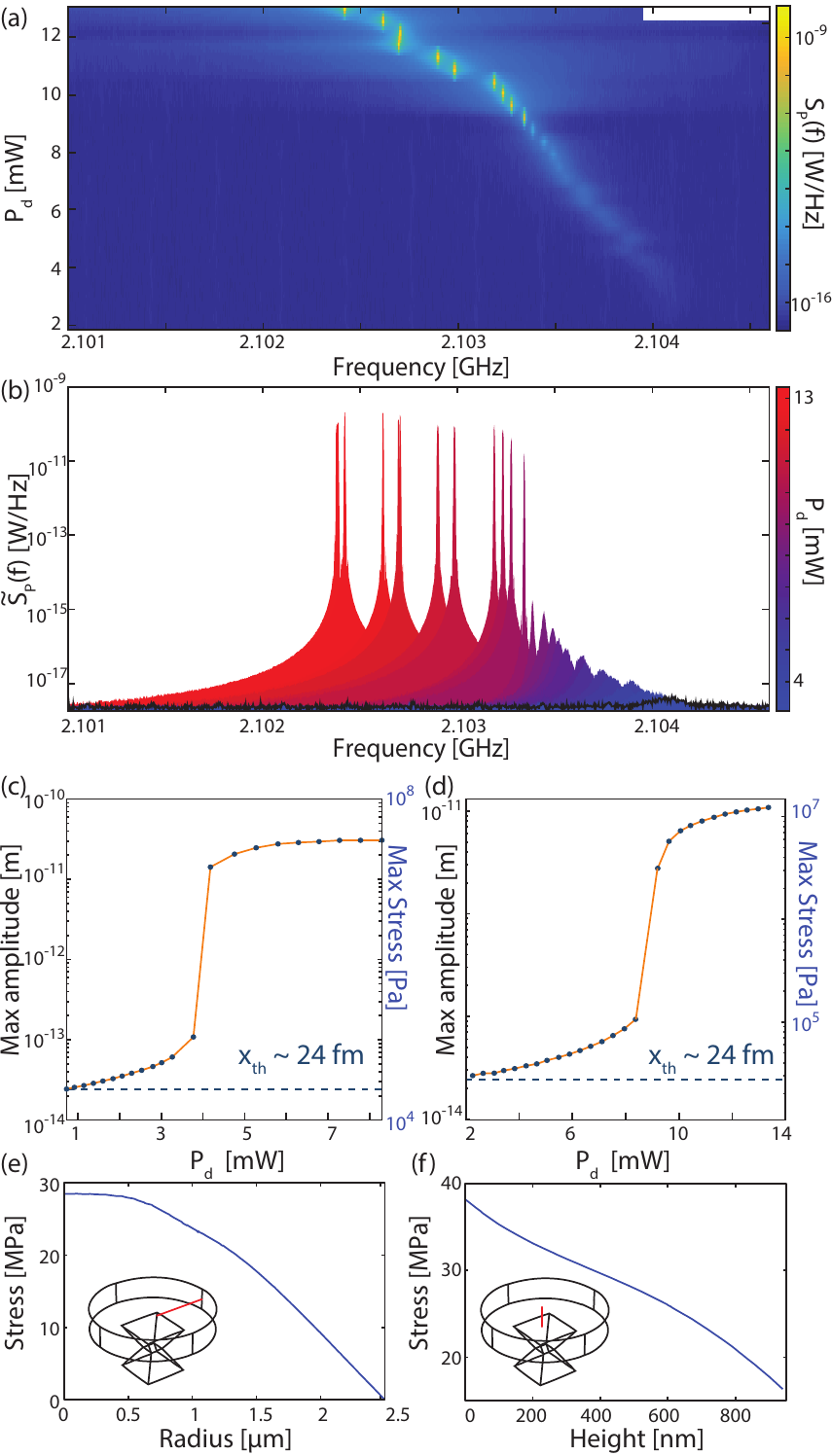, width=1\linewidth}
 \caption{Observation of microdisk self-oscillation. (a) $S_P(f;\lambda_\text{opt})$ as a function of dropped power. (b) Normalized cross-sections of (a), where $\tilde{S}_P(f)$ is given by $S_P(f;\lambda_\text{opt})$ normalized by the transduction gain so that the area under the curve represents the mechanical energy of the RBM. The black data is the thermal displacement spectra.
 (c, d) Maximum displacement amplitude and stress for $5\,\mu\text{m}$ diameter devices with (c) $Q_\text{m} \sim 9000$, $Q^{(t)}_\text{o} \sim 6 \times 10^4$, and (d) $Q_\text{m} \sim 8000 $, $Q^{(t)}_\text{o} \sim 4 \times 10^4$. (e,f) Simulated stress along (e) radial and (f) vertical cuts in the microdisk, as indicated by the red lines in the insets, for the self-oscillating amplitude in (c).}
\label{fig:self-oscillation}
\end{figure}

At higher power the microdisk optomechanical dynamics can be dramatically modified. Figure \ref{fig:self-oscillation} shows the behavior of the microdisk RBM when the input power to the fiber taper is increased sufficiently for $P_d$ to reach 13 mW.  This elevated power level corresponds to an intracavity photon number $N \sim 2.8\times 10^6$, and an optomechanical cooperativity $C = N g_0^2/\gamma_\text{o}\gamma_\text{m}= 2.7$ for the device shown in Fig.\ \ref{fig:SEM}(a). This $C$ exceeds all previously reported values for devices operating in ambient conditions \cite{ ref:safavi2011eit, ref:liu2013eit}. For such large $P_d$, if the input $\lambda_\text{s}$ is appropriately blue detuned from $\lambda_\text{o}$, it is possible for $\delta\gamma_\text{m} + \gamma_\text{m}\to 0$, resulting in self-oscillation of the microdisk RBM. Operation in this regime results in large dynamical strain within the microdisk, offering a potential path for achieving for large NV spin-phonon coupling.

We predict the strain achievable in our devices from measurements as follows. The microdisk mechanical response in the transition from thermal motion to self-oscillation is shown in Fig.\ \ref{fig:self-oscillation}(a), which displays $S_P(f;\lambda_\text{opt})$ for varying $P_d$, with $\lambda_\text{s}$ tuned to the value $\lambda_\text{opt}$ where $S_P(f_\text{m})$ is maximum. As $P_d$ is increased the mechanical resonance is observed to narrow and increase in amplitude, suggestive of the onset of self-oscillations, also referred to as phonon lasing. This is more clearly illustrated in Fig.\ \ref{fig:self-oscillation}(b), which shows the normalized spectrum $\tilde{S}_P(f)$ for varying $P_d$. Here $\tilde{S}_P(f)$ has been obtained by normalizing $S_P(f;\lambda_\text{opt})$ with the power dependent transduction gain, such that the area under $\tilde{S}_P(f)$ represents the mechanical energy of the RBM, i.e.\ $\tilde{S}_P$ is constant with respect to $P_i$ in absence of optomechanical backaction (see Supplement 1, Section 2).  At low power, this mechanical energy is dominantly from the thermal bath, and is predicted from the equipartition theorem to correspond to oscillation amplitude  $x_{\text{th}} \sim 24\,\text{fm}$.  Figures \ref{fig:self-oscillation}(c) and (d) show that for large $P_d$ a maximum $x_\text{om} = 31\,\text{pm}$ is reached, likely limited by nonlinearity of the material.  The corresponding predicted stress maximum associated with the self-oscillations, also shown in Figs.\ \ref{fig:self-oscillation}(c) and (d), is $> 30\,\text{MPa}$. Stress values were determined from $x_\text{om}$ and finite element simulations of the RBM displacement field shown in Fig.\ \ref{fig:SEM}(a), and the maximum is predicted to occur at the center the microdisk top surface, as shown by the plots in  Figs.\ \ref{fig:self-oscillation}(e,f). The corresponding maximum strain is $\approx 30\times 10^{-6}$.

The self-oscillation threshold behaviour can be quantitatively analyzed by extracting the  mechanical displacement amplitude $x_\text{om}$ as a function of $P_d$ from  $\tilde{S}_P$. This is shown Figs.\ \ref{fig:self-oscillation}(c,d) for two similarly sized microdisks. In each case, a clear threshold is observed. Since these microdisks have different $Q_\text{m}$ and $Q_\text{o}$ (see Fig.\ \ref{fig:self-oscillation} caption), their threshold power $P_T$ differ. For devices close to the sideband resolved regime, the optimal detuning for self-oscillation to occur is $\Delta \sim \omega_\text{m}$, and $P_T$ is given by  \cite{ref:wei2012hfs}
\begin{equation}\label{eq:pth}
P_T = \frac{m_\text{eff}\omega_\text{o}}{2g^2_\text{om}} \frac{\gamma_m\gamma^{(i)}_\text{o}}{\omega_m\gamma^{(t)}_\text{o}}({\gamma^{(t)}_\text{o}}/{2})^2[(2\omega_\text{m})^2+(\gamma^{(t)}_\text{o}/2)^2]
\end{equation}
\noindent where $g_\text{om} = g_0/x_\text{zpm}$ is the optomechanical coupling coefficient, and $\gamma^{(i)}_\text{o}$ and $\gamma^{(t)}_\text{o}$ are the intrinsic and fiber taper loaded optical decay rates, respectively. Here $x_\text{zpm} = \sqrt{\hbar/2 m_\text{eff}\omega_\text{m}}$  is the mechanical zero point motion amplitude. For the devices studied here, $x_\text{zpm} \sim 0.32\,\text{fm}$, as calculated from the RBM effective mass $m_\text{eff} \sim 40\,\text{pg}$, predicted by the finite element simulated displacement field shown in Fig.\ \ref{fig:SEM}(a) \cite{ref:aspelmeyer2014co}. The observed $P_T = 3.5\,\text{mW}$ and $8.5\,\text{mW}$, for the devices in  Figs.\ \ref{fig:self-oscillation}(c) and \ref{fig:self-oscillation}(d) respectively, are above the predicted values of $760\,\mu\text{W}$ and $3.0\,\text{mW}$ obtained from \eqref{eq:pth} assuming $g_0$ is given by the fits in Fig.\ \ref{fig:optomechanics}. This disagreement could be related to uncertainty in $\gamma_\text{o}$ given that $P_T$ scales to the fourth power of this quantity, interplay between doublets that is ignored by \eqref{eq:pth}, and uncertainty in $\Delta$ inferred from the cavity response in the presence of thermo-optic dispersion.

\section{Device potential for hybrid spin-optomechanics}

\noindent The potential of these devices for hybrid spin-optomechanics applications can be measured by the predicted strain coupling rate $g_{e^\text{-}}$  between a single phonon of the microdisk RBM and a single diamond NV center electron spin.  The maximum zero point motion strain of the RBM is $\epsilon_\text{zpm} \approx 3 \times 10^{-10}$, and we  estimate $g_{e^\text{-}}/2\pi= d\,\epsilon_\text{zpm} \approx 6\,\text{Hz}$ for a negatively charged NV$^-$ center electron spin optimally located $50\,\text{nm}$ below the top surface of the device in Fig.\ \ref{fig:self-oscillation}(c), exceeding the highest rate demonstrated to date \cite{ref:meesala2016esc}. Here $d \approx 10-20$ GHz is the strain susceptibility of the ground state spin \cite{ref:barfuss2015smd, ref:ovartchaiyapong2014dsc}.  When the RBM is self-oscillating as in Fig.\ \ref{fig:self-oscillation}(c), the predicted coupling rate is $G/2\pi \approx 0.6\,\text{MHz}$  \cite{ref:ovartchaiyapong2014dsc}. This is comparable to coupling rates achieved in  piezoelectric actuated nanomechanical \cite{ref:barfuss2015smd, ref:ovartchaiyapong2014dsc, ref:meesala2016esc} and bulk devices \cite{ref:macquarrie2013msc,ref:macquarrie2015ccn}. The longest room-temperature ground state spin decoherence time ($T_2$) observed to date in isotopically engineered single-crystal diamond is 1.8 ms \cite{ref:balasubramanian2009usc}, while typical $T_2$ values in nanostructures are on the order of 100 $\mu$s \cite{ref:hodges2012lln, ref:ovartchaiyapong2014dsc}. Additionally, dynamical decoupling schemes can be utilized to extend this time, as $T_2 \sim 600$ ms has been observed at low temperature \cite{ref:bar-gill2013ses}. Photon--spin control should be possible provided $G/2\pi > T_2$, which is the case for these devices.

The GHz frequency of the RBM enables low room temperature phonon occupation, relevant for cooling to the quantum ground state from room temperature \cite{ref:chan2011lcn}. This also enables access to larger energy spin transitions \cite{ref:macquarrie2013msc,ref:macquarrie2015ccn} than possible using previously demonstrated diamond nanomechanical resonators. This may be particularly important for future studies of phonon coupling to the NV$^-$ center excited state manifold, which could achieve single phonon coupling rates close to a MHz due to the $\sim 10^5$ times larger strain susceptibility of the excited states \cite{ref:davies1976oso, ref:batalov2009lts, ref:macquarrie2016cmr, ref:golter2016oqc}. This is promising for implementing fully quantum photon-phonon-spin interfaces, and for proposals of spin-mediated cooling of nanomechanical resonators \cite{ref:wilson2004lcn, ref:kepesidis2013pcl, ref:macquarrie2016cmr}.

In the samples under study, we expect to find NVs optimally coupled to the RBM since the nitrogen concentration for this diamond sample ($\sim$ ppm, corresponding to a number density of $1.76\times 10^{5} \mu\text{m}^{-3}$) results in high-concentration NV ensembles. However, future studies with higher purity samples may require NV implantation to optimally locate NVs $\sim$ 50 nm below the device surface. Additionally, due to minimal coupling of fluorescence from an NV centre located at the centre of the disk to the optical modes, free space collection would most likely be required. However, use of higher order radial breathing modes could allow for greater spatial overlap of the strain and electromagnetic field maxima \cite{ref:baker2014pcg}, allowing for more efficient fiber based excitation and collection.

Future improvements of $Q_\text{o}$ to values above $10^5$ and approaching $10^6$ should be possible  \cite{ref:burek2014hqf, ref:khanaliloo2015hqv}, enabling ultralow self-oscillation threshold  \cite{ref:wei2012hfs} and operation deep in the sideband resolved regime required for optomechanical ground state cooling \cite{ref:chan2011lcn}. Operating in vacuum, at low temperature, and using devices fabricated from high-purity electronic grade diamond may allow further increases in $Q_\text{m}$  \cite{ref:tao2013scd}, boosting the achievable photon--phonon  cooperativity $C$ and $Q_\text{m}\cdot f_\text{m}$ product. Similarly, reducing the microdisk diameter may increase $C$ through enhanced $g_0$, while also increasing $\omega_\text{m}$. Simulations predict that diameters close to $\text{3.5}~\mu\text{m}$ are possible before radiation loss limits $Q_\text{o} < 10^5$; such devices would have $g_0/2\pi > 95$ kHz and $f_\text{m} \sim 3.4\,\text{GHz}$. Finally, using electronic grade diamond material and investigating processing techniques to reduce surface state absorption may decrease optical absorption and allow larger $N$ before device heating becomes significant.

\section{Conclusion}
\noindent In conclusion, we have shown that cavity optomechanical devices can be realized from single-crystal diamond, with record high ambient condition optomecanical cooperativity, $C\sim3$, and $Q_\text{m}\cdot f_\text{m}$ product of $1.9 \times 10^{13}$. These devices are a promising testbed for ambient condition coherent optomechanics experiments, e.g.\ ground state cooling \cite{ref:chan2011lcn}, optomechanically induced transparency \cite{ref:weis2010oit, ref:safavi2011eit,ref:liu2013eit} and phonon mediated wavelength conversion \cite{ref:dong2012odm, ref:hill2012cow, ref:liu2013eit}, as well as studies in quantum information science \cite{ref:stannigel2012oqi}, and hybrid quantum systems involving light, phonons, and diamond NV center spins \cite{ref:barfuss2015smd, ref:ovartchaiyapong2014dsc, ref:macquarrie2013msc, ref:meesala2016esc, ref:ramos2013nqo, ref:wilson2004lcn, ref:kepesidis2013pcl}. We have also shown that  the microdisks demonstrated here support high-$Q_\text{o}$ optical modes at wavelengths near--resonant with the $637~\text{nm}$ optical transition of NV centers, further enhancing their potential for photon-phonon-spin coupling experiments.

We note that, in parallel to this work, Burek et al.\ have demonstrated cavity optomechanics in single--crystal diamond optomechanical crystals fabricated by a Faraday cage angled-etching technique \cite{ref:burek2015doc}.

\subsection*{Funding.}
\noindent National Research Council Canada (NRC); Natural Sciences and Engineering Research Council of Canada (NSERC); National Institute for Nanotechnology (NINT), Canada Foundation for Innovation (CFI); Alberta Innovates--Technology Futures (AITF).

\subsection*{Acknowledgment.}
\noindent The authors would like to thank A.\ C.\ Hryciw for his assistance with this project.

\input{manuscript_bibliography.bbl}

\clearpage
\normalsize
\onecolumngrid

\setcounter{equation}{0}
\setcounter{figure}{0}
\setcounter{section}{0}
\setcounter{subsection}{0}
\setcounter{table}{0}
\setcounter{page}{1}
\makeatletter
\renewcommand{\theequation}{S\arabic{equation}}
\renewcommand{\thetable}{S\arabic{table}}
\renewcommand{\thefigure}{S\arabic{figure}}
\renewcommand{ \citenumfont}[1]{S#1}
\renewcommand{\bibnumfmt}[1]{[S#1]}

\section*{Single-crystal diamond low-dissipation cavity optomechanics: supplementary material}

\section{Thermal shift and bistability}

\noindent Here we outline the process for extracting the power dependent detuning, $\Delta$. This process follows Carmon et al. \cite{ref:supp_carmon2004dtb}, beginning with the expression for the shifted cavity resonance wavelength as a function of temperature, in thermal equilibrium
\begin{align}
\lambda_\text{o}'(\Delta \text{T})&= \lambda_\text{o}+\Delta \lambda_\text{o}\,,\\
                                &= \lambda_\text{o}\left[1+\left( \eta_{\epsilon}\epsilon+\eta_{T}\frac{1}{n}\frac{d n}{d \text{T}} \right) \Delta \text{T}\right]\,,\\
                                &=\lambda_\text{o}\left[1+a \Delta \text{T}\right]\label{eq:temp_shift}.
\end{align}

\noindent This expression is obtained by considering thermal expansion of the cavity, determined by the thermal expansion coefficient $\epsilon$, and the thermo-optic effect, which shifts the refractive index $n$ with temperature $\text{T}$. Here $\eta_\text{T}$ and $\eta_\epsilon$ are geometric factors accounting for the optical mode overlap with the changing $n$ and volume, respectively. Lumped constant $a$ describes the net thermo-optic dispersion of the cavity mode. Using the room temperature single--crystal diamond values of $\epsilon \sim 1 \times 10^{-6}$ and $dn/d\text{T} \sim 1 \times 10^{-5}$ we can estimate the change in temperature of the cavity as

\begin{equation}\label{eq:thermal}
\Delta\text{T} = \left[\frac{\lambda_\text{o}'(\Delta \text{T})}{\lambda_\text{o}} - 1\right]\cdot \frac{1}{a}.
\end{equation}

\begin{figure}[h]
\epsfig{figure=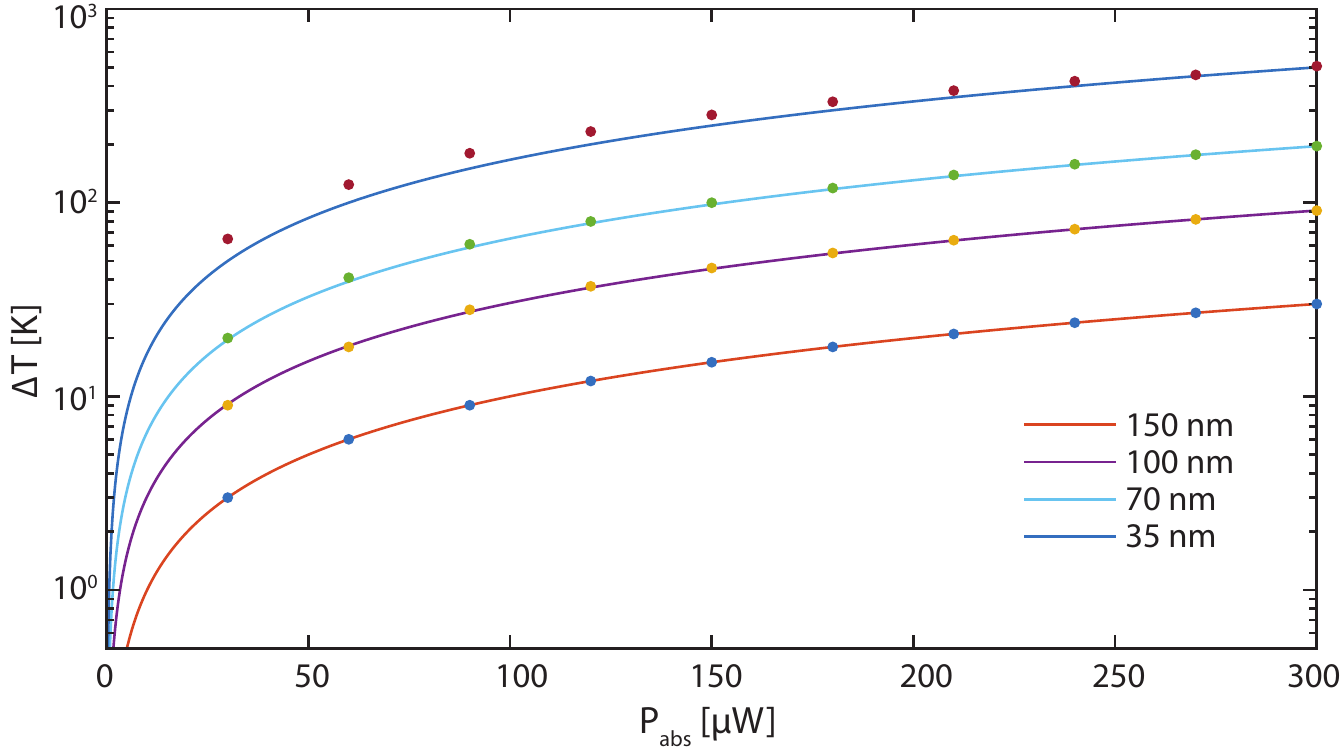, width=0.70\linewidth}
 \caption{Simulated change in temperature, $\Delta\text{T}$ of a $\sim5\,\mu$m diameter microdisk as a function of absorbed power, $P_\text{abs}$} for varying pedestal widths. Here the total heat flow to the device is given by $P_\text{abs}/V$, where $V$ is the volume defined by the outer edge of the microdisk, with $V \sim 2\,\mu\text{m}^3$. Each line represents a linear line of best fit to $\Delta\text{T}$ as a function of $P_\text{abs}$.
\label{fig:Delta_T}
\end{figure}

\noindent The shift of $\Delta \lambda_\text{o}\sim 400$ pm, as seen in Fig. 2(a) of the main text, corresponds to a change in device temperature $\Delta\text{T} \sim 50$ K. In this device the diamond forming the $\sim 100$ nm diameter pedestal has a significantly smaller thermal conductivity than that of bulk diamond ($\text{K}\sim 1500\,\text{W}\text{m}^{-1}\text{K}^{-1}$), reaching values $< 100\,\text{W}\text{m}^{-1}\text{K}^{-1}$ for nanowires $< 100$ nm in diameter \cite{ref:supp_li2012tcd}. In order to confirm that the cavity temperature shift predicted by \eqref{eq:thermal} was reasonable for our system we performed finite element COMSOL simulations to estimate $\Delta\text{T}$, including the modified thermal conductivity for the pedestal, as shown in Fig.\,\ref{fig:Delta_T} for varying pedestal widths. Fig.\,\ref{fig:Delta_T} indicates that for a pedestal width of $\sim\,100$ nm, and corresponding diamond thermal conductivity of $\sim 300\,\text{W}\text{m}^{-1}\text{K}^{-1}$ a shift of 50 K is expected when $P_\text{abs} \sim 170\,\mu$W, where $P_\text{abs}$ is the total power absorbed by the cavity. This corresponds to an optical absorption rate, $\gamma_\text{abs} \times 2\pi \sim $ 312 MHz, which is $\sim 10 \%$ of the total cavity decay rate, $\gamma_\text{tot}$. A linear relationship between $\Delta\text{T}$ and $P_\text{abs}$ is observed for the pedestal thicknesses studied here.

To convert (\ref{eq:temp_shift}) to a form that depends on the experimentally measured, normalized cavity transmission $\overline{T}$, we treat the microdisk as being in thermal equilibrium with its environment such that
\begin{equation}
\dot{q}_\text{in}=\frac{\gamma_\text{abs}}{\gamma_\text{tot}}P_d\,,
\end{equation}
where $\dot{q}_\text{in}$, and $P_d$ are the heat flow and power dropped into the cavity, respectively. Furthermore, we assume that
\begin{equation}
\dot{q}_\text{out}=K \Delta \text{T}\,,
\end{equation}
where $K$ is the thermal conductivity between
the cavity mode volume and the surrounding \cite{ref:supp_carmon2004dtb}.
In thermal equilibrium the heat flow into the cavity will be equal to the heat flow out of the cavity, which allows us to write the equilibrium temperature as
\begin{equation}
\Delta \text{T} = \frac{\gamma_\text{abs}}{\gamma_\text{tot}} \frac{P_d}{K}.
\end{equation}

\noindent Next we observe that since $P_d=(1-\overline{T})P_{i}$ where $P_{i}$ is the fiber taper waveguide input power, we can write
\begin{equation}
\Delta \text{T} = \frac{\gamma_\text{abs}}{\gamma_\text{tot}} \frac{(1-\overline{T})P_{i}}{K}\,,
\end{equation}
and the expected cavity mode shift in terms of the resonance contrast
\begin{align}
\lambda_\text{o}'(\Delta \text{T}) &=\lambda_\text{o}\left[1+a \Delta \text{T}\right]\,,\\
                                &= \lambda_\text{o}\left[1+\left(\frac{a}{K}  \frac{\gamma_\text{abs}}{\gamma_\text{tot}} P_{i}  \right) (1-\overline{T}) \right]\,,\\
                                &= \lambda_\text{o}\left[1+d                                                                                                               (1-\overline{T}) \right].
\end{align}
This gives the laser-cavity wavelength detuning, $\Delta_\lambda$  as
\begin{align}
\Delta_{\lambda} &= \lambda_\text{s}-\lambda_\text{o}'\,,\\
&=\lambda_\text{s}-\lambda_\text{o}-d (1-\overline{T}).
\end{align}
\noindent where $d = \frac{a}{K}  \frac{\gamma_{abs}}{\gamma_{tot}} P_{i}$ is used as a free parameter in fitting our cavity transmission profile. The laser detuning $\Delta$ can then be calculated for any bistable lineshape.

\section{Self Oscillations and Displacement Amplitude}

\noindent In the weak damping regime ($\gamma_\text{m} \ll \omega_\text{m}$) the oscillation amplitude of a thermally driven harmonic oscillator is given by the equipartition theorem \cite{ref:supp_aspelmeyer2014co} as
\begin{equation}
x_\text{th} = \sqrt{\frac{k_B \text{T}} {m_{\text{eff}}\omega_\text{m}^2}}\,,
\end{equation}
\noindent where $k_B$ is the Boltzmann constant, T = 295 K is the bath temperature, and $m_{\text{eff}}=$ 40 pg and $\omega_\text{m}/2\pi \sim 2$ GHz are the effective mass and mechanical frequency of the radial breathing mode studied here, respectively. This results in $x_\text{th} = 24$ fm and a zero point fluctuation motion, $x_\text{zpm} = 0.32$ fm.

While $S_P(f)\propto \langle x^2\rangle$, where $\langle x^2\rangle$ is the variance of the mechanical displacement, one must be more careful when calculating the mechanical energy. Strictly speaking $\langle x^2\rangle$ is related to the single sided displacement spectral density $S_{xx}(\omega)$ by
\begin{equation}
\langle x^2\rangle = \int^{\infty}_{0}S_{xx}(\omega)\frac{d\omega}{2\pi}.
\end{equation}
\noindent This can be connected to the measured cavity transmission noise spectrum $S_P(\omega)$ through a cavity transfer function $H(\omega,\Delta)$, $P_i$, and $g_\text{om}$ \cite{ref:supp_lin2010cmo}
\begin{equation}
S_P(\omega) = g_\text{om}^2P_i^2S_{xx}(\omega)H(\omega,\Delta).
\end{equation}
\noindent In this experiment we measure $S_P(\omega)$, and can compute the area under the curve, $A$, given by $A = \int^{\infty}_{0}S_P(\omega)\frac{d\omega}{2\pi}$.  If we change $P_i$ from $P_{i_1}$ to $P_{i_2}$, keep $\Delta$ constant,and ignore the small ($\sim 0.02 \%$) changes in $\omega_\text{m}$, such that $H(\omega, \Delta; P_{i_1}) = H(\omega, \Delta; P_{i_2})$,  we can show that the ratio of the area under the curve corresponding to $P_{i_1}$ and $P_{i_2}$ given by $A_1$ and $A_2$, respectively, is
\begin{equation}
\frac{A_1}{A_2} = \frac{P_{i_1}^2\langle x_1^2\rangle}{P_{i_2}^2\langle x_2^2\rangle}.
\end{equation}
\noindent We can then calibrate high $P_{i}$ measurements to the thermal case, where $P_i$ is small enough for optomechanical backaction effects to be ignored.  The displacement amplitude, $x_\text{om}$, of the RBM in the self-oscillation regime can then be calculated as
\begin{equation}
x_{\text{om}} =  x_{\text{th}}   \sqrt{\frac{A_\text{om}}{A_\text{th}}\frac{P_T^2}{P_\text{om}^2} }\,,
\end{equation}
\noindent where $\text{A}_\text{om}$ and $\text{A}_\text{th}$ are the area under the curve in the driven ($P_i = P_\text{om}$) and thermal  ($P_i = P_T$) states, respectively.  Similarly, for the purpose of comparing mechanical spectra it is useful to calculate the normalized cavity transmission noise spectrum $\tilde{S}_P$, given by
\begin{equation}
\tilde{S}_P(\omega; P_i, \Delta) = \left. S_P(\omega; P_i) \frac{P_i^2}{P_T^2}\right|_\Delta.
\end{equation}

The maximum oscillation amplitude $x_\text{om}$ is shown as a function of dropped optical power in Fig.\ 3(a), where the absolute maximum oscillation amplitude was found to be $\sim$ 31 pm ($\sim x_{\text{th}} \cdot 10^3$ ). Using finite element COMSOL simulations and by assuming that diamond behaves as a linear elastic material in the self oscillation regime, these amplitudes correspond to stress on the order of tens of MPa at the center of the microdisk.

\section{Comparison of $Q_\text{m}\cdot \lowercase{f}_\text{m}$ product}

\begin{figure*}[h]
\begin{center}
\epsfig{figure=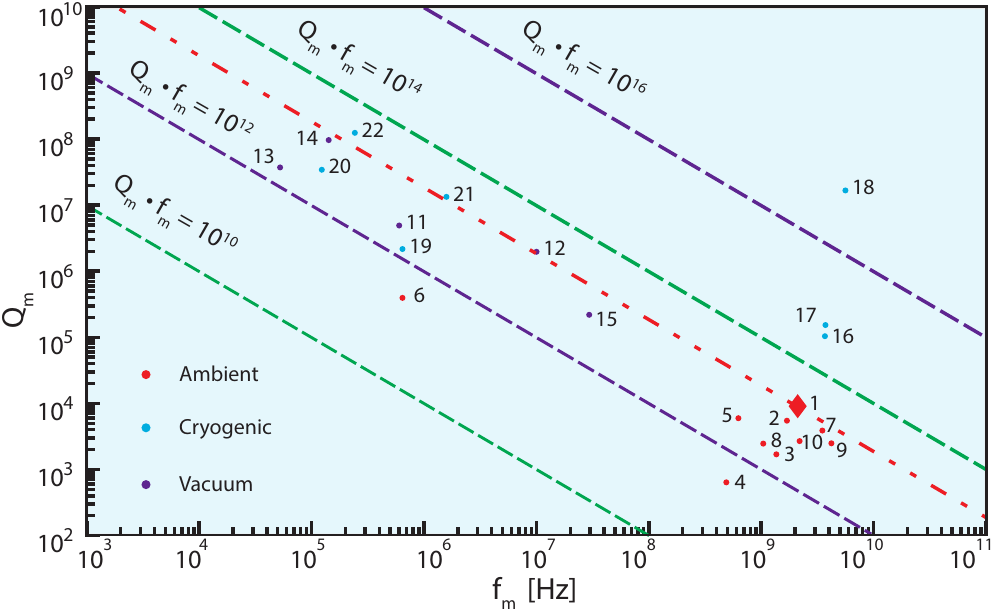, width=0.70\linewidth}
 \caption{Comparison of high $Q_\text{m}\cdot f_\text{m}$ product products for a variety of optomechanical systems, as listed in Table \ref{table:comp}.}
\label{fig:Qf}
\end{center}
\end{figure*}

\noindent The device studied here demonstrates the largest $Q_\text{m}\cdot f_\text{m}$ product of an optomechanical device measured in ambient conditions to date. Figure\ \ref{fig:Qf} compares this value with a survey of some of the largest $Q_\text{m}\cdot f_\text{m}$ products observed in cavity optomechanical systems in ambient, cryogenic, and low pressure conditions. Note that higher $Q_\text{m}\cdot f_\text{m}$ products have been demonstrated compared to this work, but required either vacuum or low-temperature environments.

\vspace{-5mm}

\begin{table}[h]
\small
\centering
\caption{\textbf{Survey of highest $Q_\text{m}\cdot f_\text{m}$ products observed in cavity optomechanical systems to date, corresponding to those shown in Fig.\ \ref{fig:Qf}.}}
\begin{tabular}{cccc}
  \hline
  No. & Author/Reference & Material & Structure  \\ \hline
  1 & Mitchell et al. (This Work) & Diamond & Microdisk  \\
  2 & Lu et al. \cite{ref:supp_lu2015hfh} & SiC & Microdisk   \\
  3 & Nguyen et al. \cite{ref:supp_nguyen2013uqf} & GaAs & Microdisk  \\
  4 & Mitchell et al. \cite{ref:supp_mitchell2014cog} & GaP & Microdisk  \\
  5 & Liu et al. \cite{ref:supp_liu2013eit} & Si$_3$N$_4$ & Microdisk   \\
  6 & Fong et al. \cite{ref:supp_fong2012fpn} & Si$_3$N$_4$ & Beam \& Waveguide   \\
  7 & Grutter et al. \cite{ref:supp_grutter2015soc} & Si$_3$N$_4$ & Optomechanical Crystal   \\
  8 & Xiong et al. \cite{ref:supp_xiong2012ihf} & AlN & Suspended Ring Resonator   \\
  9 & Bochmann et al. \cite{ref:supp_bochmann2013ncb} & AlN & Optomechanical Crystal   \\
  10 & Eichenfield et al. \cite{ref:supp_eichenfield2009oc} & Si & Optomechanical Crystal  \\ \hline
  11 & Bui et al. \cite{ref:supp_bui2012hrh} & Si$_3$N$_4$ & Membrane Photonic Crystal + Fabry P$\acute{\text{e}}$rot Cavity   \\
  12 & Wilson et al. \cite{wilson2009cos} & Si$_3$N$_4$ & Membrane + Fabry P$\acute{\text{e}}$rot Cavity  \\
  13 & Reinhardt et al. \cite{ref:supp_reinhardt2016uns} & Si$_3$N$_4$ & Membrane + Fabry P$\acute{\text{e}}$rot Cavity \\
  14 & Norte et al. \cite{ref:supp_norte2016mrq} & Si$_3$N$_4$ & Membrane Photonic Crystal + Fabry P$\acute{\text{e}}$rot Cavity \\
  15 & Zhang et al. \cite{ref:supp_zhang2015itf} & Si$_3$N$_4$ & Tuning Fork  + Microdisk \\ \hline
  16 & Chan et al.\cite{ref:supp_chan2011lcn} & Si & Optomechanical Crystal + Phononic Shield    \\
  17 & Krause et al. \cite{ref:supp_krause2015nrp} & Si & Optomechanical Crystal + Phononic Shield   \\
  18 & Meenehan et al. \cite{ref:supp_meenehan2015ped} & Si & Optomechanical Crystal + Phononic Shield  \\
  19 & Fong et al. \cite{ref:supp_fong2012fpn} & Si$_3$N$_4$ & Beam + On-chip Interferometer \\
  20 & Yuan et al. \cite{ref:supp_yuan2015lcm} & Si$_3$N$_4$ & Membrane + Superconducting Microwave Cavity    \\
  21 & Purdy et al. \cite{ref:supp_purdy2012cos} & Si$_3$N$_4$ & Membrane + Fabry P$\acute{\text{e}}$rot Cavity  \\
  22 & Yuan et al. \cite{ref:supp_yuan2015snm} & Si$_3$N$_4$ & Membrane + Superconducting Microwave Cavity   \\
  \hline
\end{tabular}
\label{table:comp}
\end{table}

\newpage

\input{supplementary_bibliography.bbl}
\end{document}

%% file: manuscript_bibliography.bbl
%

%% file: SCD_low_dissipation_cavity_optomechanics_arXiv.bbl
\begin{thebibliography}{65}%
\makeatletter
\providecommand \@ifxundefined [1]{%
 \@ifx{#1\undefined}
}%
\providecommand \@ifnum [1]{%
 \ifnum #1\expandafter \@firstoftwo
 \else \expandafter \@secondoftwo
 \fi
}%
\providecommand \@ifx [1]{%
 \ifx #1\expandafter \@firstoftwo
 \else \expandafter \@secondoftwo
 \fi
}%
\providecommand \natexlab [1]{#1}%
\providecommand \enquote  [1]{``#1''}%
\providecommand \bibnamefont  [1]{#1}%
\providecommand \bibfnamefont [1]{#1}%
\providecommand \citenamefont [1]{#1}%
\providecommand \href@noop [0]{\@secondoftwo}%
\providecommand \href [0]{\begingroup \@sanitize@url \@href}%
\providecommand \@href[1]{\@@startlink{#1}\@@href}%
\providecommand \@@href[1]{\endgroup#1\@@endlink}%
\providecommand \@sanitize@url [0]{\catcode `\\12\catcode `\$12\catcode
  `\&12\catcode `\#12\catcode `\^12\catcode `\_12\catcode `\%12\relax}%
\providecommand \@@startlink[1]{}%
\providecommand \@@endlink[0]{}%
\providecommand \url  [0]{\begingroup\@sanitize@url \@url }%
\providecommand \@url [1]{\endgroup\@href {#1}{\urlprefix }}%
\providecommand \urlprefix  [0]{URL }%
\providecommand \Eprint [0]{\href }%
\providecommand \doibase [0]{http://dx.doi.org/}%
\providecommand \selectlanguage [0]{\@gobble}%
\providecommand \bibinfo  [0]{\@secondoftwo}%
\providecommand \bibfield  [0]{\@secondoftwo}%
\providecommand \translation [1]{[#1]}%
\providecommand \BibitemOpen [0]{}%
\providecommand \bibitemStop [0]{}%
\providecommand \bibitemNoStop [0]{.\EOS\space}%
\providecommand \EOS [0]{\spacefactor3000\relax}%
\providecommand \BibitemShut  [1]{\csname bibitem#1\endcsname}%
\let\auto@bib@innerbib\@empty
\bibitem [{\citenamefont {Treutlein}\ \emph {et~al.}(2014)\citenamefont
  {Treutlein}, \citenamefont {Genes}, \citenamefont {Hammerer}, \citenamefont
  {Poggio},\ and\ \citenamefont {Rabl}}]{ref:treutlein2014hms}%
  \BibitemOpen
  \bibfield  {author} {\bibinfo {author} {\bibfnamefont {P.}~\bibnamefont
  {Treutlein}}, \bibinfo {author} {\bibfnamefont {C.}~\bibnamefont {Genes}},
  \bibinfo {author} {\bibfnamefont {K.}~\bibnamefont {Hammerer}}, \bibinfo
  {author} {\bibfnamefont {M.}~\bibnamefont {Poggio}}, \ and\ \bibinfo {author}
  {\bibfnamefont {P.}~\bibnamefont {Rabl}},\ }in\ \href@noop {} {\emph
  {\bibinfo {booktitle} {Cavity Optomechanics}}}\ (\bibinfo  {publisher}
  {Springer},\ \bibinfo {year} {2014})\ pp.\ \bibinfo {pages}
  {327--351}\BibitemShut {NoStop}%
\bibitem [{\citenamefont {Aharonovich}\ \emph {et~al.}(2011)\citenamefont
  {Aharonovich}, \citenamefont {Greentree},\ and\ \citenamefont
  {Prawer}}]{ref:aharonovich2011dp}%
  \BibitemOpen
  \bibfield  {author} {\bibinfo {author} {\bibfnamefont {I.}~\bibnamefont
  {Aharonovich}}, \bibinfo {author} {\bibfnamefont {A.~D.}\ \bibnamefont
  {Greentree}}, \ and\ \bibinfo {author} {\bibfnamefont {S.}~\bibnamefont
  {Prawer}},\ }\href@noop {} {\bibfield  {journal} {\bibinfo  {journal} {Nature
  Photon.}\ }\textbf {\bibinfo {volume} {5}},\ \bibinfo {pages} {397} (\bibinfo
  {year} {2011})}\BibitemShut {NoStop}%
\bibitem [{\citenamefont {MacQuarrie}\ \emph {et~al.}(2013)\citenamefont
  {MacQuarrie}, \citenamefont {Gosavi}, \citenamefont {Jungwirth},
  \citenamefont {Bhave},\ and\ \citenamefont {Fuchs}}]{ref:macquarrie2013msc}%
  \BibitemOpen
  \bibfield  {author} {\bibinfo {author} {\bibfnamefont {E.~R.}\ \bibnamefont
  {MacQuarrie}}, \bibinfo {author} {\bibfnamefont {T.~A.}\ \bibnamefont
  {Gosavi}}, \bibinfo {author} {\bibfnamefont {N.~R.}\ \bibnamefont
  {Jungwirth}}, \bibinfo {author} {\bibfnamefont {S.~A.}\ \bibnamefont
  {Bhave}}, \ and\ \bibinfo {author} {\bibfnamefont {G.~D.}\ \bibnamefont
  {Fuchs}},\ }\href@noop {} {\bibfield  {journal} {\bibinfo  {journal} {Phys.
  Rev. Lett.}\ }\textbf {\bibinfo {volume} {111}},\ \bibinfo {pages} {227602}
  (\bibinfo {year} {2013})}\BibitemShut {NoStop}%
\bibitem [{\citenamefont {Teissier}\ \emph {et~al.}(2014)\citenamefont
  {Teissier}, \citenamefont {Barfuss}, \citenamefont {Appel}, \citenamefont
  {Neu},\ and\ \citenamefont {Maletinsky}}]{ref:teissier2014scn}%
  \BibitemOpen
  \bibfield  {author} {\bibinfo {author} {\bibfnamefont {J.}~\bibnamefont
  {Teissier}}, \bibinfo {author} {\bibfnamefont {A.}~\bibnamefont {Barfuss}},
  \bibinfo {author} {\bibfnamefont {P.}~\bibnamefont {Appel}}, \bibinfo
  {author} {\bibfnamefont {E.}~\bibnamefont {Neu}}, \ and\ \bibinfo {author}
  {\bibfnamefont {P.}~\bibnamefont {Maletinsky}},\ }\href@noop {} {\bibfield
  {journal} {\bibinfo  {journal} {Phys. Rev. Lett.}\ }\textbf {\bibinfo
  {volume} {113}},\ \bibinfo {pages} {020503} (\bibinfo {year}
  {2014})}\BibitemShut {NoStop}%
\bibitem [{\citenamefont {Ovartchaiyapong}\ \emph {et~al.}(2014)\citenamefont
  {Ovartchaiyapong}, \citenamefont {Lee}, \citenamefont {Myers},\ and\
  \citenamefont {Jayich}}]{ref:ovartchaiyapong2014dsc}%
  \BibitemOpen
  \bibfield  {author} {\bibinfo {author} {\bibfnamefont {P.}~\bibnamefont
  {Ovartchaiyapong}}, \bibinfo {author} {\bibfnamefont {K.~W.}\ \bibnamefont
  {Lee}}, \bibinfo {author} {\bibfnamefont {B.~A.}\ \bibnamefont {Myers}}, \
  and\ \bibinfo {author} {\bibfnamefont {A.~C.~B.}\ \bibnamefont {Jayich}},\
  }\href@noop {} {\bibfield  {journal} {\bibinfo  {journal} {Nat. Commun.}\
  }\textbf {\bibinfo {volume} {5}},\ \bibinfo {pages} {4429} (\bibinfo {year}
  {2014})}\BibitemShut {NoStop}%
\bibitem [{\citenamefont {Barfuss}\ \emph {et~al.}(2015)\citenamefont
  {Barfuss}, \citenamefont {Teissier}, \citenamefont {Neu}, \citenamefont
  {Nunnenkamp},\ and\ \citenamefont {Maletinsky}}]{ref:barfuss2015smd}%
  \BibitemOpen
  \bibfield  {author} {\bibinfo {author} {\bibfnamefont {A.}~\bibnamefont
  {Barfuss}}, \bibinfo {author} {\bibfnamefont {J.}~\bibnamefont {Teissier}},
  \bibinfo {author} {\bibfnamefont {E.}~\bibnamefont {Neu}}, \bibinfo {author}
  {\bibfnamefont {A.}~\bibnamefont {Nunnenkamp}}, \ and\ \bibinfo {author}
  {\bibfnamefont {P.}~\bibnamefont {Maletinsky}},\ }\href@noop {} {\bibfield
  {journal} {\bibinfo  {journal} {Nature Phys.}\ }\textbf {\bibinfo {volume}
  {11}},\ \bibinfo {pages} {820} (\bibinfo {year} {2015})}\BibitemShut
  {NoStop}%
\bibitem [{\citenamefont {Meesala}\ \emph {et~al.}(2016)\citenamefont
  {Meesala}, \citenamefont {Sohn}, \citenamefont {Atikian}, \citenamefont
  {Kim}, \citenamefont {Burek}, \citenamefont {Choy},\ and\ \citenamefont
  {Lon\ifmmode~\check{c}\else \v{c}\fi{}ar}}]{ref:meesala2016esc}%
  \BibitemOpen
  \bibfield  {author} {\bibinfo {author} {\bibfnamefont {S.}~\bibnamefont
  {Meesala}}, \bibinfo {author} {\bibfnamefont {Y.-I.}\ \bibnamefont {Sohn}},
  \bibinfo {author} {\bibfnamefont {H.~A.}\ \bibnamefont {Atikian}}, \bibinfo
  {author} {\bibfnamefont {S.}~\bibnamefont {Kim}}, \bibinfo {author}
  {\bibfnamefont {M.~J.}\ \bibnamefont {Burek}}, \bibinfo {author}
  {\bibfnamefont {J.~T.}\ \bibnamefont {Choy}}, \ and\ \bibinfo {author}
  {\bibfnamefont {M.}~\bibnamefont {Lon\ifmmode~\check{c}\else \v{c}\fi{}ar}},\
  }\href {\doibase 10.1103/PhysRevApplied.5.034010} {\bibfield  {journal}
  {\bibinfo  {journal} {Phys. Rev. Applied}\ }\textbf {\bibinfo {volume} {5}},\
  \bibinfo {pages} {034010} (\bibinfo {year} {2016})}\BibitemShut {NoStop}%
\bibitem [{\citenamefont {MacQuarrie}\ \emph {et~al.}(2015)\citenamefont
  {MacQuarrie}, \citenamefont {Gosavi}, \citenamefont {Moehle}, \citenamefont
  {Jungwirth}, \citenamefont {Bhave},\ and\ \citenamefont
  {Fuchs}}]{ref:macquarrie2015ccn}%
  \BibitemOpen
  \bibfield  {author} {\bibinfo {author} {\bibfnamefont {E.~R.}\ \bibnamefont
  {MacQuarrie}}, \bibinfo {author} {\bibfnamefont {T.~A.}\ \bibnamefont
  {Gosavi}}, \bibinfo {author} {\bibfnamefont {A.~M.}\ \bibnamefont {Moehle}},
  \bibinfo {author} {\bibfnamefont {N.~R.}\ \bibnamefont {Jungwirth}}, \bibinfo
  {author} {\bibfnamefont {S.~A.}\ \bibnamefont {Bhave}}, \ and\ \bibinfo
  {author} {\bibfnamefont {G.~D.}\ \bibnamefont {Fuchs}},\ }\href {\doibase
  10.1364/OPTICA.2.000233} {\bibfield  {journal} {\bibinfo  {journal} {Optica}\
  }\textbf {\bibinfo {volume} {2}},\ \bibinfo {pages} {233} (\bibinfo {year}
  {2015})}\BibitemShut {NoStop}%
\bibitem [{\citenamefont {Arcizet}\ \emph {et~al.}(2011)\citenamefont
  {Arcizet}, \citenamefont {Jacques}, \citenamefont {Siria}, \citenamefont
  {Poncharal}, \citenamefont {Vincent},\ and\ \citenamefont
  {Seidelin}}]{ref:arcizet2011snv}%
  \BibitemOpen
  \bibfield  {author} {\bibinfo {author} {\bibfnamefont {O.}~\bibnamefont
  {Arcizet}}, \bibinfo {author} {\bibfnamefont {V.}~\bibnamefont {Jacques}},
  \bibinfo {author} {\bibfnamefont {A.}~\bibnamefont {Siria}}, \bibinfo
  {author} {\bibfnamefont {P.}~\bibnamefont {Poncharal}}, \bibinfo {author}
  {\bibfnamefont {P.}~\bibnamefont {Vincent}}, \ and\ \bibinfo {author}
  {\bibfnamefont {S.}~\bibnamefont {Seidelin}},\ }\href@noop {} {\bibfield
  {journal} {\bibinfo  {journal} {Nature Phys.}\ }\textbf {\bibinfo {volume}
  {7}},\ \bibinfo {pages} {879} (\bibinfo {year} {2011})}\BibitemShut {NoStop}%
\bibitem [{\citenamefont {Golter}\ \emph {et~al.}(2016)\citenamefont {Golter},
  \citenamefont {Oo}, \citenamefont {Amezcua}, \citenamefont {Stewart},\ and\
  \citenamefont {Wang}}]{ref:golter2016oqc}%
  \BibitemOpen
  \bibfield  {author} {\bibinfo {author} {\bibfnamefont {D.~A.}\ \bibnamefont
  {Golter}}, \bibinfo {author} {\bibfnamefont {T.}~\bibnamefont {Oo}}, \bibinfo
  {author} {\bibfnamefont {M.}~\bibnamefont {Amezcua}}, \bibinfo {author}
  {\bibfnamefont {K.~A.}\ \bibnamefont {Stewart}}, \ and\ \bibinfo {author}
  {\bibfnamefont {H.}~\bibnamefont {Wang}},\ }\href {\doibase
  10.1103/PhysRevLett.116.143602} {\bibfield  {journal} {\bibinfo  {journal}
  {Phys. Rev. Lett.}\ }\textbf {\bibinfo {volume} {116}},\ \bibinfo {pages}
  {143602} (\bibinfo {year} {2016})}\BibitemShut {NoStop}%
\bibitem [{\citenamefont {Lee}\ \emph {et~al.}(2016)\citenamefont {Lee},
  \citenamefont {Lee}, \citenamefont {Ovartchaiyapong}, \citenamefont
  {Minguzzi}, \citenamefont {Maze},\ and\ \citenamefont
  {Jayich}}]{ref:lee2016scm}%
  \BibitemOpen
  \bibfield  {author} {\bibinfo {author} {\bibfnamefont {K.~W.}\ \bibnamefont
  {Lee}}, \bibinfo {author} {\bibfnamefont {D.}~\bibnamefont {Lee}}, \bibinfo
  {author} {\bibfnamefont {P.}~\bibnamefont {Ovartchaiyapong}}, \bibinfo
  {author} {\bibfnamefont {J.}~\bibnamefont {Minguzzi}}, \bibinfo {author}
  {\bibfnamefont {J.~R.}\ \bibnamefont {Maze}}, \ and\ \bibinfo {author}
  {\bibfnamefont {A.~C.~B.}\ \bibnamefont {Jayich}},\ }\href@noop {} {\bibfield
   {journal} {\bibinfo  {journal} {arXiv:13603.07680}\ } (\bibinfo {year}
  {2016})}\BibitemShut {NoStop}%
\bibitem [{\citenamefont {Aspelmeyer}\ \emph {et~al.}(2014)\citenamefont
  {Aspelmeyer}, \citenamefont {Kippenberg},\ and\ \citenamefont
  {Marquardt}}]{ref:aspelmeyer2014co}%
  \BibitemOpen
  \bibfield  {author} {\bibinfo {author} {\bibfnamefont {M.}~\bibnamefont
  {Aspelmeyer}}, \bibinfo {author} {\bibfnamefont {T.~J.}\ \bibnamefont
  {Kippenberg}}, \ and\ \bibinfo {author} {\bibfnamefont {F.}~\bibnamefont
  {Marquardt}},\ }\href {\doibase 10.1103/RevModPhys.86.1391} {\bibfield
  {journal} {\bibinfo  {journal} {Rev. Mod. Phys.}\ }\textbf {\bibinfo {volume}
  {86}},\ \bibinfo {pages} {1391} (\bibinfo {year} {2014})}\BibitemShut
  {NoStop}%
\bibitem [{\citenamefont {Weis}\ \emph {et~al.}(2010)\citenamefont {Weis},
  \citenamefont {Rivi{\`e}re}, \citenamefont {Del{\'e}glise}, \citenamefont
  {Gavartin}, \citenamefont {Arcizet}, \citenamefont {Schliesser},\ and\
  \citenamefont {Kippenberg}}]{ref:weis2010oit}%
  \BibitemOpen
  \bibfield  {author} {\bibinfo {author} {\bibfnamefont {S.}~\bibnamefont
  {Weis}}, \bibinfo {author} {\bibfnamefont {R.}~\bibnamefont {Rivi{\`e}re}},
  \bibinfo {author} {\bibfnamefont {S.}~\bibnamefont {Del{\'e}glise}}, \bibinfo
  {author} {\bibfnamefont {E.}~\bibnamefont {Gavartin}}, \bibinfo {author}
  {\bibfnamefont {O.}~\bibnamefont {Arcizet}}, \bibinfo {author} {\bibfnamefont
  {A.}~\bibnamefont {Schliesser}}, \ and\ \bibinfo {author} {\bibfnamefont
  {T.~J.}\ \bibnamefont {Kippenberg}},\ }\href@noop {} {\bibfield  {journal}
  {\bibinfo  {journal} {Science}\ }\textbf {\bibinfo {volume} {330}},\ \bibinfo
  {pages} {1520} (\bibinfo {year} {2010})}\BibitemShut {NoStop}%
\bibitem [{\citenamefont {Safavi-Naeini}\ \emph {et~al.}(2011)\citenamefont
  {Safavi-Naeini}, \citenamefont {Alegre}, \citenamefont {Chan}, \citenamefont
  {Eichenfield}, \citenamefont {Winger}, \citenamefont {Lin}, \citenamefont
  {Hill}, \citenamefont {Chang},\ and\ \citenamefont
  {Painter}}]{ref:safavi2011eit}%
  \BibitemOpen
  \bibfield  {author} {\bibinfo {author} {\bibfnamefont {A.~H.}\ \bibnamefont
  {Safavi-Naeini}}, \bibinfo {author} {\bibfnamefont {T.~M.}\ \bibnamefont
  {Alegre}}, \bibinfo {author} {\bibfnamefont {J.}~\bibnamefont {Chan}},
  \bibinfo {author} {\bibfnamefont {M.}~\bibnamefont {Eichenfield}}, \bibinfo
  {author} {\bibfnamefont {M.}~\bibnamefont {Winger}}, \bibinfo {author}
  {\bibfnamefont {Q.}~\bibnamefont {Lin}}, \bibinfo {author} {\bibfnamefont
  {J.~T.}\ \bibnamefont {Hill}}, \bibinfo {author} {\bibfnamefont
  {D.}~\bibnamefont {Chang}}, \ and\ \bibinfo {author} {\bibfnamefont
  {O.}~\bibnamefont {Painter}},\ }\href@noop {} {\bibfield  {journal} {\bibinfo
   {journal} {Nature}\ }\textbf {\bibinfo {volume} {472}},\ \bibinfo {pages}
  {69} (\bibinfo {year} {2011})}\BibitemShut {NoStop}%
\bibitem [{\citenamefont {Liu}\ \emph {et~al.}(2013)\citenamefont {Liu},
  \citenamefont {Davan{\c{c}}o}, \citenamefont {Aksyuk},\ and\ \citenamefont
  {Srinivasan}}]{ref:liu2013eit}%
  \BibitemOpen
  \bibfield  {author} {\bibinfo {author} {\bibfnamefont {Y.}~\bibnamefont
  {Liu}}, \bibinfo {author} {\bibfnamefont {M.}~\bibnamefont {Davan{\c{c}}o}},
  \bibinfo {author} {\bibfnamefont {V.}~\bibnamefont {Aksyuk}}, \ and\ \bibinfo
  {author} {\bibfnamefont {K.}~\bibnamefont {Srinivasan}},\ }\href@noop {}
  {\bibfield  {journal} {\bibinfo  {journal} {Phys. Rev. Lett.}\ }\textbf
  {\bibinfo {volume} {110}},\ \bibinfo {pages} {223603} (\bibinfo {year}
  {2013})}\BibitemShut {NoStop}%
\bibitem [{\citenamefont {Chan}\ \emph {et~al.}(2011)\citenamefont {Chan},
  \citenamefont {Alegre}, \citenamefont {Safavi-Naeini}, \citenamefont {Hill},
  \citenamefont {Krause}, \citenamefont {Groblacher}, \citenamefont
  {Aspelmeyer},\ and\ \citenamefont {Painter}}]{ref:chan2011lcn}%
  \BibitemOpen
  \bibfield  {author} {\bibinfo {author} {\bibfnamefont {J.}~\bibnamefont
  {Chan}}, \bibinfo {author} {\bibfnamefont {T.~P.~M.}\ \bibnamefont {Alegre}},
  \bibinfo {author} {\bibfnamefont {A.~H.}\ \bibnamefont {Safavi-Naeini}},
  \bibinfo {author} {\bibfnamefont {J.~T.}\ \bibnamefont {Hill}}, \bibinfo
  {author} {\bibfnamefont {A.}~\bibnamefont {Krause}}, \bibinfo {author}
  {\bibfnamefont {S.}~\bibnamefont {Groblacher}}, \bibinfo {author}
  {\bibfnamefont {M.}~\bibnamefont {Aspelmeyer}}, \ and\ \bibinfo {author}
  {\bibfnamefont {O.}~\bibnamefont {Painter}},\ }\href@noop {} {\bibfield
  {journal} {\bibinfo  {journal} {Nature}\ }\textbf {\bibinfo {volume} {478}},\
  \bibinfo {pages} {89} (\bibinfo {year} {2011})}\BibitemShut {NoStop}%
\bibitem [{\citenamefont {Kepesidis}\ \emph {et~al.}(2013)\citenamefont
  {Kepesidis}, \citenamefont {Bennett}, \citenamefont {Portolan}, \citenamefont
  {Lukin},\ and\ \citenamefont {Rabl}}]{ref:kepesidis2013pcl}%
  \BibitemOpen
  \bibfield  {author} {\bibinfo {author} {\bibfnamefont {K.~V.}\ \bibnamefont
  {Kepesidis}}, \bibinfo {author} {\bibfnamefont {S.~D.}\ \bibnamefont
  {Bennett}}, \bibinfo {author} {\bibfnamefont {S.}~\bibnamefont {Portolan}},
  \bibinfo {author} {\bibfnamefont {M.~D.}\ \bibnamefont {Lukin}}, \ and\
  \bibinfo {author} {\bibfnamefont {P.}~\bibnamefont {Rabl}},\ }\href@noop {}
  {\bibfield  {journal} {\bibinfo  {journal} {Phys. Rev. B}\ }\textbf {\bibinfo
  {volume} {88}},\ \bibinfo {pages} {064105} (\bibinfo {year}
  {2013})}\BibitemShut {NoStop}%
\bibitem [{\citenamefont {Rabl}\ \emph {et~al.}(2010)\citenamefont {Rabl},
  \citenamefont {Kolkowitz}, \citenamefont {Koppens}, \citenamefont {Harris},
  \citenamefont {Zoller},\ and\ \citenamefont {Lukin}}]{ref:rabl2010qst}%
  \BibitemOpen
  \bibfield  {author} {\bibinfo {author} {\bibfnamefont {P.}~\bibnamefont
  {Rabl}}, \bibinfo {author} {\bibfnamefont {S.}~\bibnamefont {Kolkowitz}},
  \bibinfo {author} {\bibfnamefont {F.}~\bibnamefont {Koppens}}, \bibinfo
  {author} {\bibfnamefont {J.}~\bibnamefont {Harris}}, \bibinfo {author}
  {\bibfnamefont {P.}~\bibnamefont {Zoller}}, \ and\ \bibinfo {author}
  {\bibfnamefont {M.}~\bibnamefont {Lukin}},\ }\href@noop {} {\bibfield
  {journal} {\bibinfo  {journal} {Nature Phys.}\ }\textbf {\bibinfo {volume}
  {6}},\ \bibinfo {pages} {602} (\bibinfo {year} {2010})}\BibitemShut {NoStop}%
\bibitem [{\citenamefont {Stannigel}\ \emph {et~al.}(2010)\citenamefont
  {Stannigel}, \citenamefont {Rabl}, \citenamefont {S{\o}rensen}, \citenamefont
  {Zoller},\ and\ \citenamefont {Lukin}}]{ref:stannigel2010otl}%
  \BibitemOpen
  \bibfield  {author} {\bibinfo {author} {\bibfnamefont {K.}~\bibnamefont
  {Stannigel}}, \bibinfo {author} {\bibfnamefont {P.}~\bibnamefont {Rabl}},
  \bibinfo {author} {\bibfnamefont {A.~S.}\ \bibnamefont {S{\o}rensen}},
  \bibinfo {author} {\bibfnamefont {P.}~\bibnamefont {Zoller}}, \ and\ \bibinfo
  {author} {\bibfnamefont {M.~D.}\ \bibnamefont {Lukin}},\ }\href@noop {}
  {\bibfield  {journal} {\bibinfo  {journal} {Phys. Rev. Lett.}\ }\textbf
  {\bibinfo {volume} {105}},\ \bibinfo {pages} {220501} (\bibinfo {year}
  {2010})}\BibitemShut {NoStop}%
\bibitem [{\citenamefont {Schuetz}\ \emph {et~al.}(2015)\citenamefont
  {Schuetz}, \citenamefont {Kessler}, \citenamefont {Giedke}, \citenamefont
  {Vandersypen}, \citenamefont {Lukin},\ and\ \citenamefont
  {Cirac}}]{ref:schuetz2015uqt}%
  \BibitemOpen
  \bibfield  {author} {\bibinfo {author} {\bibfnamefont {M.~J.~A.}\
  \bibnamefont {Schuetz}}, \bibinfo {author} {\bibfnamefont {E.~M.}\
  \bibnamefont {Kessler}}, \bibinfo {author} {\bibfnamefont {G.}~\bibnamefont
  {Giedke}}, \bibinfo {author} {\bibfnamefont {L.~M.~K.}\ \bibnamefont
  {Vandersypen}}, \bibinfo {author} {\bibfnamefont {M.~D.}\ \bibnamefont
  {Lukin}}, \ and\ \bibinfo {author} {\bibfnamefont {J.~I.}\ \bibnamefont
  {Cirac}},\ }\href {\doibase 10.1103/PhysRevX.5.031031} {\bibfield  {journal}
  {\bibinfo  {journal} {Phys. Rev. X}\ }\textbf {\bibinfo {volume} {5}},\
  \bibinfo {pages} {031031} (\bibinfo {year} {2015})}\BibitemShut {NoStop}%
\bibitem [{\citenamefont {Wilson-Rae}\ \emph {et~al.}(2004)\citenamefont
  {Wilson-Rae}, \citenamefont {Zoller},\ and\ \citenamefont
  {Imamo\v{g}lu}}]{ref:wilson2004lcn}%
  \BibitemOpen
  \bibfield  {author} {\bibinfo {author} {\bibfnamefont {I.}~\bibnamefont
  {Wilson-Rae}}, \bibinfo {author} {\bibfnamefont {P.}~\bibnamefont {Zoller}},
  \ and\ \bibinfo {author} {\bibfnamefont {A.}~\bibnamefont {Imamo\v{g}lu}},\
  }\href@noop {} {\bibfield  {journal} {\bibinfo  {journal} {Phys. Rev. Lett.}\
  }\textbf {\bibinfo {volume} {92}},\ \bibinfo {pages} {075507} (\bibinfo
  {year} {2004})}\BibitemShut {NoStop}%
\bibitem [{\citenamefont {MacQuarrie}\ \emph {et~al.}(2016)\citenamefont
  {MacQuarrie}, \citenamefont {Otten}, \citenamefont {Gray},\ and\
  \citenamefont {Fuchs}}]{ref:macquarrie2016cmr}%
  \BibitemOpen
  \bibfield  {author} {\bibinfo {author} {\bibfnamefont {E.}~\bibnamefont
  {MacQuarrie}}, \bibinfo {author} {\bibfnamefont {M.}~\bibnamefont {Otten}},
  \bibinfo {author} {\bibfnamefont {S.}~\bibnamefont {Gray}}, \ and\ \bibinfo
  {author} {\bibfnamefont {G.}~\bibnamefont {Fuchs}},\ }\href@noop {}
  {\bibfield  {journal} {\bibinfo  {journal} {arXiv:1605.07131}\ } (\bibinfo
  {year} {2016})}\BibitemShut {NoStop}%
\bibitem [{\citenamefont {Ramos}\ \emph {et~al.}(2013)\citenamefont {Ramos},
  \citenamefont {Sudhir}, \citenamefont {Stannigel}, \citenamefont {Zoller},\
  and\ \citenamefont {Kippenberg}}]{ref:ramos2013nqo}%
  \BibitemOpen
  \bibfield  {author} {\bibinfo {author} {\bibfnamefont {T.}~\bibnamefont
  {Ramos}}, \bibinfo {author} {\bibfnamefont {V.}~\bibnamefont {Sudhir}},
  \bibinfo {author} {\bibfnamefont {K.}~\bibnamefont {Stannigel}}, \bibinfo
  {author} {\bibfnamefont {P.}~\bibnamefont {Zoller}}, \ and\ \bibinfo {author}
  {\bibfnamefont {T.~J.}\ \bibnamefont {Kippenberg}},\ }\href@noop {}
  {\bibfield  {journal} {\bibinfo  {journal} {Phys. Rev. Lett.}\ }\textbf
  {\bibinfo {volume} {110}},\ \bibinfo {pages} {193602} (\bibinfo {year}
  {2013})}\BibitemShut {NoStop}%
\bibitem [{\citenamefont {Rath}\ \emph {et~al.}(2013)\citenamefont {Rath},
  \citenamefont {Khasminskaya}, \citenamefont {Nebel}, \citenamefont {Wild},\
  and\ \citenamefont {Pernice}}]{ref:rath2013dio}%
  \BibitemOpen
  \bibfield  {author} {\bibinfo {author} {\bibfnamefont {P.}~\bibnamefont
  {Rath}}, \bibinfo {author} {\bibfnamefont {S.}~\bibnamefont {Khasminskaya}},
  \bibinfo {author} {\bibfnamefont {C.}~\bibnamefont {Nebel}}, \bibinfo
  {author} {\bibfnamefont {C.}~\bibnamefont {Wild}}, \ and\ \bibinfo {author}
  {\bibfnamefont {W.~H.}\ \bibnamefont {Pernice}},\ }\href@noop {} {\bibfield
  {journal} {\bibinfo  {journal} {Nat. Commun.}\ }\textbf {\bibinfo {volume}
  {4}},\ \bibinfo {pages} {1690} (\bibinfo {year} {2013})}\BibitemShut
  {NoStop}%
\bibitem [{\citenamefont {Tao}\ \emph {et~al.}(2013)\citenamefont {Tao},
  \citenamefont {Boss}, \citenamefont {Moores},\ and\ \citenamefont
  {Degen}}]{ref:tao2013scd}%
  \BibitemOpen
  \bibfield  {author} {\bibinfo {author} {\bibfnamefont {Y.}~\bibnamefont
  {Tao}}, \bibinfo {author} {\bibfnamefont {J.~M.}\ \bibnamefont {Boss}},
  \bibinfo {author} {\bibfnamefont {B.~A.}\ \bibnamefont {Moores}}, \ and\
  \bibinfo {author} {\bibfnamefont {C.~L.}\ \bibnamefont {Degen}},\ }\href@noop
  {} {\bibfield  {journal} {\bibinfo  {journal} {Nat. Commun.}\ }\textbf
  {\bibinfo {volume} {5}},\ \bibinfo {pages} {3638} (\bibinfo {year}
  {2013})}\BibitemShut {NoStop}%
\bibitem [{\citenamefont {Balasubramanian}\ \emph {et~al.}(2009)\citenamefont
  {Balasubramanian}, \citenamefont {Neumann}, \citenamefont {Twitchen},
  \citenamefont {Markham}, \citenamefont {Kolesov}, \citenamefont {Mizuochi},
  \citenamefont {Isoya}, \citenamefont {Achard}, \citenamefont {Beck},
  \citenamefont {Tissler}, \citenamefont {Jacques}, \citenamefont {Hemmer},
  \citenamefont {Jelezko},\ and\ \citenamefont
  {Wrachtrup}}]{ref:balasubramanian2009usc}%
  \BibitemOpen
  \bibfield  {author} {\bibinfo {author} {\bibfnamefont {G.}~\bibnamefont
  {Balasubramanian}}, \bibinfo {author} {\bibfnamefont {P.}~\bibnamefont
  {Neumann}}, \bibinfo {author} {\bibfnamefont {D.}~\bibnamefont {Twitchen}},
  \bibinfo {author} {\bibfnamefont {M.}~\bibnamefont {Markham}}, \bibinfo
  {author} {\bibfnamefont {R.}~\bibnamefont {Kolesov}}, \bibinfo {author}
  {\bibfnamefont {N.}~\bibnamefont {Mizuochi}}, \bibinfo {author}
  {\bibfnamefont {J.}~\bibnamefont {Isoya}}, \bibinfo {author} {\bibfnamefont
  {J.}~\bibnamefont {Achard}}, \bibinfo {author} {\bibfnamefont
  {J.}~\bibnamefont {Beck}}, \bibinfo {author} {\bibfnamefont {J.}~\bibnamefont
  {Tissler}}, \bibinfo {author} {\bibfnamefont {V.}~\bibnamefont {Jacques}},
  \bibinfo {author} {\bibfnamefont {P.}~\bibnamefont {Hemmer}}, \bibinfo
  {author} {\bibfnamefont {F.}~\bibnamefont {Jelezko}}, \ and\ \bibinfo
  {author} {\bibfnamefont {F.}~\bibnamefont {Wrachtrup}},\ }\href@noop {}
  {\bibfield  {journal} {\bibinfo  {journal} {Nature Mater.}\ }\textbf
  {\bibinfo {volume} {8}},\ \bibinfo {pages} {383} (\bibinfo {year}
  {2009})}\BibitemShut {NoStop}%
\bibitem [{\citenamefont {Gil-Santos}\ \emph {et~al.}(2015)\citenamefont
  {Gil-Santos}, \citenamefont {Baker}, \citenamefont {Nguyen}, \citenamefont
  {Hease}, \citenamefont {Lema{\^\i}tre}, \citenamefont {Ducci}, \citenamefont
  {Leo},\ and\ \citenamefont {Favero}}]{ref:gil2015hfn}%
  \BibitemOpen
  \bibfield  {author} {\bibinfo {author} {\bibfnamefont {E.}~\bibnamefont
  {Gil-Santos}}, \bibinfo {author} {\bibfnamefont {C.}~\bibnamefont {Baker}},
  \bibinfo {author} {\bibfnamefont {D.}~\bibnamefont {Nguyen}}, \bibinfo
  {author} {\bibfnamefont {W.}~\bibnamefont {Hease}}, \bibinfo {author}
  {\bibfnamefont {A.}~\bibnamefont {Lema{\^\i}tre}}, \bibinfo {author}
  {\bibfnamefont {S.}~\bibnamefont {Ducci}}, \bibinfo {author} {\bibfnamefont
  {G.}~\bibnamefont {Leo}}, \ and\ \bibinfo {author} {\bibfnamefont
  {I.}~\bibnamefont {Favero}},\ }\href@noop {} {\bibfield  {journal} {\bibinfo
  {journal} {Nature Nanotech.}\ }\textbf {\bibinfo {volume} {10}},\ \bibinfo
  {pages} {810–816} (\bibinfo {year} {2015})}\BibitemShut {NoStop}%
\bibitem [{\citenamefont {Lu}\ \emph {et~al.}(2015)\citenamefont {Lu},
  \citenamefont {Lee},\ and\ \citenamefont {Lin}}]{ref:lu2015hfh}%
  \BibitemOpen
  \bibfield  {author} {\bibinfo {author} {\bibfnamefont {X.}~\bibnamefont
  {Lu}}, \bibinfo {author} {\bibfnamefont {J.~Y.}\ \bibnamefont {Lee}}, \ and\
  \bibinfo {author} {\bibfnamefont {Q.}~\bibnamefont {Lin}},\ }\href@noop {}
  {\bibfield  {journal} {\bibinfo  {journal} {Sci. Rep.}\ }\textbf {\bibinfo
  {volume} {5}},\ \bibinfo {pages} {17005} (\bibinfo {year}
  {2015})}\BibitemShut {NoStop}%
\bibitem [{\citenamefont {Faraon}\ \emph {et~al.}(2011)\citenamefont {Faraon},
  \citenamefont {Barclay}, \citenamefont {Santori}, \citenamefont {Fu},\ and\
  \citenamefont {Beausoleil}}]{ref:faraon2011rez}%
  \BibitemOpen
  \bibfield  {author} {\bibinfo {author} {\bibfnamefont {A.}~\bibnamefont
  {Faraon}}, \bibinfo {author} {\bibfnamefont {P.~E.}\ \bibnamefont {Barclay}},
  \bibinfo {author} {\bibfnamefont {C.}~\bibnamefont {Santori}}, \bibinfo
  {author} {\bibfnamefont {K.-M.~C.}\ \bibnamefont {Fu}}, \ and\ \bibinfo
  {author} {\bibfnamefont {R.~G.}\ \bibnamefont {Beausoleil}},\ }\href@noop {}
  {\bibfield  {journal} {\bibinfo  {journal} {Nature Photon.}\ }\textbf
  {\bibinfo {volume} {5}},\ \bibinfo {pages} {301} (\bibinfo {year}
  {2011})}\BibitemShut {NoStop}%
\bibitem [{\citenamefont {Ovartchaiyapong}\ \emph {et~al.}(2012)\citenamefont
  {Ovartchaiyapong}, \citenamefont {Pascal}, \citenamefont {Myers},
  \citenamefont {Lauria},\ and\ \citenamefont
  {Bleszynski~Jayich}}]{ref:ovartchaiyapong2012hqf}%
  \BibitemOpen
  \bibfield  {author} {\bibinfo {author} {\bibfnamefont {P.}~\bibnamefont
  {Ovartchaiyapong}}, \bibinfo {author} {\bibfnamefont {L.~M.~A.}\ \bibnamefont
  {Pascal}}, \bibinfo {author} {\bibfnamefont {B.~A.}\ \bibnamefont {Myers}},
  \bibinfo {author} {\bibfnamefont {P.}~\bibnamefont {Lauria}}, \ and\ \bibinfo
  {author} {\bibfnamefont {A.~C.}\ \bibnamefont {Bleszynski~Jayich}},\
  }\href@noop {} {\bibfield  {journal} {\bibinfo  {journal} {Appl. Phys.
  Lett.}\ }\textbf {\bibinfo {volume} {101}},\ \bibinfo {eid} {163505}
  (\bibinfo {year} {2012})}\BibitemShut {NoStop}%
\bibitem [{\citenamefont {Burek}\ \emph {et~al.}(2012)\citenamefont {Burek},
  \citenamefont {de~Leon}, \citenamefont {Shields}, \citenamefont {Hausmann},
  \citenamefont {Chu}, \citenamefont {Quan}, \citenamefont {Zibrov},
  \citenamefont {Park}, \citenamefont {Lukin},\ and\ \citenamefont
  {Lon\v{c}ar}}]{ref:burek2012fsm}%
  \BibitemOpen
  \bibfield  {author} {\bibinfo {author} {\bibfnamefont {M.~J.}\ \bibnamefont
  {Burek}}, \bibinfo {author} {\bibfnamefont {N.~P.}\ \bibnamefont {de~Leon}},
  \bibinfo {author} {\bibfnamefont {B.~J.}\ \bibnamefont {Shields}}, \bibinfo
  {author} {\bibfnamefont {B.~J.}\ \bibnamefont {Hausmann}}, \bibinfo {author}
  {\bibfnamefont {Y.}~\bibnamefont {Chu}}, \bibinfo {author} {\bibfnamefont
  {Q.}~\bibnamefont {Quan}}, \bibinfo {author} {\bibfnamefont {A.~S.}\
  \bibnamefont {Zibrov}}, \bibinfo {author} {\bibfnamefont {H.}~\bibnamefont
  {Park}}, \bibinfo {author} {\bibfnamefont {M.~D.}\ \bibnamefont {Lukin}}, \
  and\ \bibinfo {author} {\bibfnamefont {M.}~\bibnamefont {Lon\v{c}ar}},\
  }\href@noop {} {\bibfield  {journal} {\bibinfo  {journal} {Nano Lett.}\
  }\textbf {\bibinfo {volume} {12}},\ \bibinfo {pages} {6084} (\bibinfo {year}
  {2012})}\BibitemShut {NoStop}%
\bibitem [{\citenamefont {Khanaliloo}\ \emph
  {et~al.}(2015{\natexlab{a}})\citenamefont {Khanaliloo}, \citenamefont
  {Jayakumar}, \citenamefont {Hryciw}, \citenamefont {Lake}, \citenamefont
  {Kaviani},\ and\ \citenamefont {Barclay}}]{ref:khanaliloo2015dnw}%
  \BibitemOpen
  \bibfield  {author} {\bibinfo {author} {\bibfnamefont {B.}~\bibnamefont
  {Khanaliloo}}, \bibinfo {author} {\bibfnamefont {H.}~\bibnamefont
  {Jayakumar}}, \bibinfo {author} {\bibfnamefont {A.~C.}\ \bibnamefont
  {Hryciw}}, \bibinfo {author} {\bibfnamefont {D.~P.}\ \bibnamefont {Lake}},
  \bibinfo {author} {\bibfnamefont {H.}~\bibnamefont {Kaviani}}, \ and\
  \bibinfo {author} {\bibfnamefont {P.~E.}\ \bibnamefont {Barclay}},\ }\href
  {\doibase 10.1103/PhysRevX.5.041051} {\bibfield  {journal} {\bibinfo
  {journal} {Phys. Rev. X}\ }\textbf {\bibinfo {volume} {5}},\ \bibinfo {pages}
  {041051} (\bibinfo {year} {2015}{\natexlab{a}})}\BibitemShut {NoStop}%
\bibitem [{\citenamefont {Burek}\ \emph {et~al.}(2014)\citenamefont {Burek},
  \citenamefont {Chu}, \citenamefont {Liddy}, \citenamefont {Patel},
  \citenamefont {Rochman}, \citenamefont {Meesala}, \citenamefont {Hong},
  \citenamefont {Quan}, \citenamefont {Lukin},\ and\ \citenamefont
  {Lon\v{c}ar}}]{ref:burek2014hqf}%
  \BibitemOpen
  \bibfield  {author} {\bibinfo {author} {\bibfnamefont {M.~J.}\ \bibnamefont
  {Burek}}, \bibinfo {author} {\bibfnamefont {Y.}~\bibnamefont {Chu}}, \bibinfo
  {author} {\bibfnamefont {M.~S.~Z.}\ \bibnamefont {Liddy}}, \bibinfo {author}
  {\bibfnamefont {P.}~\bibnamefont {Patel}}, \bibinfo {author} {\bibfnamefont
  {J.}~\bibnamefont {Rochman}}, \bibinfo {author} {\bibfnamefont
  {S.}~\bibnamefont {Meesala}}, \bibinfo {author} {\bibfnamefont
  {W.}~\bibnamefont {Hong}}, \bibinfo {author} {\bibfnamefont {Q.}~\bibnamefont
  {Quan}}, \bibinfo {author} {\bibfnamefont {M.~D.}\ \bibnamefont {Lukin}}, \
  and\ \bibinfo {author} {\bibfnamefont {M.}~\bibnamefont {Lon\v{c}ar}},\
  }\href@noop {} {\bibfield  {journal} {\bibinfo  {journal} {Nat. Commun.}\
  }\textbf {\bibinfo {volume} {5}},\ \bibinfo {pages} {5718} (\bibinfo {year}
  {2014})}\BibitemShut {NoStop}%
\bibitem [{\citenamefont {Khanaliloo}\ \emph
  {et~al.}(2015{\natexlab{b}})\citenamefont {Khanaliloo}, \citenamefont
  {Mitchell}, \citenamefont {Hryciw},\ and\ \citenamefont
  {Barclay}}]{ref:khanaliloo2015hqv}%
  \BibitemOpen
  \bibfield  {author} {\bibinfo {author} {\bibfnamefont {B.}~\bibnamefont
  {Khanaliloo}}, \bibinfo {author} {\bibfnamefont {M.}~\bibnamefont
  {Mitchell}}, \bibinfo {author} {\bibfnamefont {A.~C.}\ \bibnamefont
  {Hryciw}}, \ and\ \bibinfo {author} {\bibfnamefont {P.~E.}\ \bibnamefont
  {Barclay}},\ }\href@noop {} {\bibfield  {journal} {\bibinfo  {journal} {Nano
  Lett.}\ }\textbf {\bibinfo {volume} {15}},\ \bibinfo {pages} {5131} (\bibinfo
  {year} {2015}{\natexlab{b}})}\BibitemShut {NoStop}%
\bibitem [{\citenamefont {Maletinsky}\ \emph {et~al.}(2012)\citenamefont
  {Maletinsky}, \citenamefont {Hong}, \citenamefont {Grinolds}, \citenamefont
  {Hausmann}, \citenamefont {Lukin}, \citenamefont {Walsworth}, \citenamefont
  {Loncar},\ and\ \citenamefont {Yacoby}}]{ref:maletinsky2012ars}%
  \BibitemOpen
  \bibfield  {author} {\bibinfo {author} {\bibfnamefont {P.}~\bibnamefont
  {Maletinsky}}, \bibinfo {author} {\bibfnamefont {S.}~\bibnamefont {Hong}},
  \bibinfo {author} {\bibfnamefont {M.~S.}\ \bibnamefont {Grinolds}}, \bibinfo
  {author} {\bibfnamefont {B.}~\bibnamefont {Hausmann}}, \bibinfo {author}
  {\bibfnamefont {M.~D.}\ \bibnamefont {Lukin}}, \bibinfo {author}
  {\bibfnamefont {R.~L.}\ \bibnamefont {Walsworth}}, \bibinfo {author}
  {\bibfnamefont {M.}~\bibnamefont {Loncar}}, \ and\ \bibinfo {author}
  {\bibfnamefont {A.}~\bibnamefont {Yacoby}},\ }\href {\doibase
  10.1038/nnano.2012.50} {\bibfield  {journal} {\bibinfo  {journal} {Nature
  Nanotech.}\ }\textbf {\bibinfo {volume} {7}},\ \bibinfo {pages} {320}
  (\bibinfo {year} {2012})}\BibitemShut {NoStop}%
\bibitem [{\citenamefont {Michael}\ \emph {et~al.}(2007)\citenamefont
  {Michael}, \citenamefont {Borselli}, \citenamefont {Johnson}, \citenamefont
  {Chrystala},\ and\ \citenamefont {Painter}}]{ref:michael2007oft}%
  \BibitemOpen
  \bibfield  {author} {\bibinfo {author} {\bibfnamefont {C.~P.}\ \bibnamefont
  {Michael}}, \bibinfo {author} {\bibfnamefont {M.}~\bibnamefont {Borselli}},
  \bibinfo {author} {\bibfnamefont {T.~J.}\ \bibnamefont {Johnson}}, \bibinfo
  {author} {\bibfnamefont {C.}~\bibnamefont {Chrystala}}, \ and\ \bibinfo
  {author} {\bibfnamefont {O.}~\bibnamefont {Painter}},\ }\href@noop {}
  {\bibfield  {journal} {\bibinfo  {journal} {Opt. Express}\ }\textbf {\bibinfo
  {volume} {15}},\ \bibinfo {pages} {4745} (\bibinfo {year}
  {2007})}\BibitemShut {NoStop}%
\bibitem [{\citenamefont {Mitchell}\ \emph {et~al.}(2014)\citenamefont
  {Mitchell}, \citenamefont {Hryciw},\ and\ \citenamefont
  {Barclay}}]{ref:mitchell2014cog}%
  \BibitemOpen
  \bibfield  {author} {\bibinfo {author} {\bibfnamefont {M.}~\bibnamefont
  {Mitchell}}, \bibinfo {author} {\bibfnamefont {A.~C.}\ \bibnamefont
  {Hryciw}}, \ and\ \bibinfo {author} {\bibfnamefont {P.~E.}\ \bibnamefont
  {Barclay}},\ }\href
  {http://scitation.aip.org/content/aip/journal/apl/104/14/10.1063/1.4870999}
  {\bibfield  {journal} {\bibinfo  {journal} {Appl. Phys. Lett.}\ }\textbf
  {\bibinfo {volume} {104}},\ \bibinfo {eid} {141104} (\bibinfo {year}
  {2014})}\BibitemShut {NoStop}%
\bibitem [{\citenamefont {Borselli}\ \emph {et~al.}(2004)\citenamefont
  {Borselli}, \citenamefont {Srinivasan}, \citenamefont {Barclay},\ and\
  \citenamefont {Painter}}]{ref:borselli2004rsm}%
  \BibitemOpen
  \bibfield  {author} {\bibinfo {author} {\bibfnamefont {M.}~\bibnamefont
  {Borselli}}, \bibinfo {author} {\bibfnamefont {K.}~\bibnamefont
  {Srinivasan}}, \bibinfo {author} {\bibfnamefont {P.~E.}\ \bibnamefont
  {Barclay}}, \ and\ \bibinfo {author} {\bibfnamefont {O.}~\bibnamefont
  {Painter}},\ }\href@noop {} {\bibfield  {journal} {\bibinfo  {journal} {Appl.
  Phys. Lett.}\ }\textbf {\bibinfo {volume} {85}},\ \bibinfo {pages} {3693}
  (\bibinfo {year} {2004})}\BibitemShut {NoStop}%
\bibitem [{\citenamefont {Nguyen}\ \emph {et~al.}(2013)\citenamefont {Nguyen},
  \citenamefont {Baker}, \citenamefont {Hease}, \citenamefont {Sejil},
  \citenamefont {Senellart}, \citenamefont {Lema\^{i}tre}, \citenamefont
  {Ducci}, \citenamefont {Leo},\ and\ \citenamefont
  {Favero}}]{ref:nguyen2013uqf}%
  \BibitemOpen
  \bibfield  {author} {\bibinfo {author} {\bibfnamefont {D.~T.}\ \bibnamefont
  {Nguyen}}, \bibinfo {author} {\bibfnamefont {C.}~\bibnamefont {Baker}},
  \bibinfo {author} {\bibfnamefont {W.}~\bibnamefont {Hease}}, \bibinfo
  {author} {\bibfnamefont {S.}~\bibnamefont {Sejil}}, \bibinfo {author}
  {\bibfnamefont {P.}~\bibnamefont {Senellart}}, \bibinfo {author}
  {\bibfnamefont {A.}~\bibnamefont {Lema\^{i}tre}}, \bibinfo {author}
  {\bibfnamefont {S.}~\bibnamefont {Ducci}}, \bibinfo {author} {\bibfnamefont
  {G.}~\bibnamefont {Leo}}, \ and\ \bibinfo {author} {\bibfnamefont
  {I.}~\bibnamefont {Favero}},\ }\href@noop {} {\bibfield  {journal} {\bibinfo
  {journal} {Appl. Phys. Lett.}\ }\textbf {\bibinfo {volume} {103}},\ \bibinfo
  {eid} {241112} (\bibinfo {year} {2013})}\BibitemShut {NoStop}%
\bibitem [{\citenamefont {Rivi{\`e}re}(2011)}]{ref:riviere:2011cos}%
  \BibitemOpen
  \bibfield  {author} {\bibinfo {author} {\bibfnamefont {R.}~\bibnamefont
  {Rivi{\`e}re}},\ }\emph {\bibinfo {title} {Cavity optomechanics with silica
  toroidal microresonators down to low phonon occupancy}},\ \href@noop {}
  {Ph.D. thesis},\ \bibinfo  {school} {Ludwig-Maximilians-Universit{\"a}t
  M{\"u}nchen} (\bibinfo {year} {2011})\BibitemShut {NoStop}%
\bibitem [{\citenamefont {Sun}\ \emph {et~al.}(2012)\citenamefont {Sun},
  \citenamefont {Zhang},\ and\ \citenamefont {Tang}}]{ref:tang2012hqs}%
  \BibitemOpen
  \bibfield  {author} {\bibinfo {author} {\bibfnamefont {X.}~\bibnamefont
  {Sun}}, \bibinfo {author} {\bibfnamefont {X.}~\bibnamefont {Zhang}}, \ and\
  \bibinfo {author} {\bibfnamefont {H.~X.}\ \bibnamefont {Tang}},\ }\href
  {\doibase http://dx.doi.org/10.1063/1.4709416} {\bibfield  {journal}
  {\bibinfo  {journal} {Appl. Phys. Lett.}\ }\textbf {\bibinfo {volume}
  {100}},\ \bibinfo {eid} {173116} (\bibinfo {year} {2012})}\BibitemShut
  {NoStop}%
\bibitem [{\citenamefont {Eichenfield}\ \emph {et~al.}(2009)\citenamefont
  {Eichenfield}, \citenamefont {Chan}, \citenamefont {Camacho}, \citenamefont
  {Vahala},\ and\ \citenamefont {Painter}}]{ref:eichenfield2009oc}%
  \BibitemOpen
  \bibfield  {author} {\bibinfo {author} {\bibfnamefont {M.}~\bibnamefont
  {Eichenfield}}, \bibinfo {author} {\bibfnamefont {J.}~\bibnamefont {Chan}},
  \bibinfo {author} {\bibfnamefont {R.}~\bibnamefont {Camacho}}, \bibinfo
  {author} {\bibfnamefont {K.}~\bibnamefont {Vahala}}, \ and\ \bibinfo {author}
  {\bibfnamefont {O.}~\bibnamefont {Painter}},\ }\href@noop {} {\bibfield
  {journal} {\bibinfo  {journal} {Nature}\ }\textbf {\bibinfo {volume} {462}},\
  \bibinfo {pages} {78} (\bibinfo {year} {2009})}\BibitemShut {NoStop}%
\bibitem [{\citenamefont {Ekinci}\ \emph {et~al.}(2004)\citenamefont {Ekinci},
  \citenamefont {Yang},\ and\ \citenamefont {Roukes}}]{ref:ekinci2004uli}%
  \BibitemOpen
  \bibfield  {author} {\bibinfo {author} {\bibfnamefont {K.}~\bibnamefont
  {Ekinci}}, \bibinfo {author} {\bibfnamefont {Y.}~\bibnamefont {Yang}}, \ and\
  \bibinfo {author} {\bibfnamefont {M.}~\bibnamefont {Roukes}},\ }\href@noop {}
  {\bibfield  {journal} {\bibinfo  {journal} {J. Appl. Phys.}\ }\textbf
  {\bibinfo {volume} {95}},\ \bibinfo {pages} {2682} (\bibinfo {year}
  {2004})}\BibitemShut {NoStop}%
\bibitem [{\citenamefont {Yu}\ \emph {et~al.}(2015)\citenamefont {Yu},
  \citenamefont {Jiang}, \citenamefont {Lin},\ and\ \citenamefont
  {Lu}}]{ref:yu2015cot}%
  \BibitemOpen
  \bibfield  {author} {\bibinfo {author} {\bibfnamefont {W.}~\bibnamefont
  {Yu}}, \bibinfo {author} {\bibfnamefont {W.~C.}\ \bibnamefont {Jiang}},
  \bibinfo {author} {\bibfnamefont {Q.}~\bibnamefont {Lin}}, \ and\ \bibinfo
  {author} {\bibfnamefont {T.}~\bibnamefont {Lu}},\ }\href@noop {} {\bibfield
  {journal} {\bibinfo  {journal} {arXiv:1504.03727}\ } (\bibinfo {year}
  {2015})}\BibitemShut {NoStop}%
\bibitem [{\citenamefont {Nguyen}(2007)}]{ref:nguyen2007mtt}%
  \BibitemOpen
  \bibfield  {author} {\bibinfo {author} {\bibfnamefont {C.~T.-C.}\
  \bibnamefont {Nguyen}},\ }\href@noop {} {\bibfield  {journal} {\bibinfo
  {journal} {IEEE Transactions on Ultrasonics, Ferroelectrics and Frequency
  Control}\ }\textbf {\bibinfo {volume} {54}},\ \bibinfo {pages} {251}
  (\bibinfo {year} {2007})}\BibitemShut {NoStop}%
\bibitem [{\citenamefont {Norte}\ \emph {et~al.}(2016)\citenamefont {Norte},
  \citenamefont {Moura},\ and\ \citenamefont
  {Gr\"oblacher}}]{ref:norte2016mrq}%
  \BibitemOpen
  \bibfield  {author} {\bibinfo {author} {\bibfnamefont {R.~A.}\ \bibnamefont
  {Norte}}, \bibinfo {author} {\bibfnamefont {J.~P.}\ \bibnamefont {Moura}}, \
  and\ \bibinfo {author} {\bibfnamefont {S.}~\bibnamefont {Gr\"oblacher}},\
  }\href {\doibase 10.1103/PhysRevLett.116.147202} {\bibfield  {journal}
  {\bibinfo  {journal} {Phys. Rev. Lett.}\ }\textbf {\bibinfo {volume} {116}},\
  \bibinfo {pages} {147202} (\bibinfo {year} {2016})}\BibitemShut {NoStop}%
\bibitem [{\citenamefont {Li}\ \emph {et~al.}(2012)\citenamefont {Li},
  \citenamefont {Mingo}, \citenamefont {Lindsay}, \citenamefont {Broido},
  \citenamefont {Stewart},\ and\ \citenamefont {Katcho}}]{ref:li2012tcd}%
  \BibitemOpen
  \bibfield  {author} {\bibinfo {author} {\bibfnamefont {W.}~\bibnamefont
  {Li}}, \bibinfo {author} {\bibfnamefont {N.}~\bibnamefont {Mingo}}, \bibinfo
  {author} {\bibfnamefont {L.}~\bibnamefont {Lindsay}}, \bibinfo {author}
  {\bibfnamefont {D.~A.}\ \bibnamefont {Broido}}, \bibinfo {author}
  {\bibfnamefont {D.~A.}\ \bibnamefont {Stewart}}, \ and\ \bibinfo {author}
  {\bibfnamefont {N.~A.}\ \bibnamefont {Katcho}},\ }\href {\doibase
  10.1103/PhysRevB.85.195436} {\bibfield  {journal} {\bibinfo  {journal} {Phys.
  Rev. B}\ }\textbf {\bibinfo {volume} {85}},\ \bibinfo {pages} {195436}
  (\bibinfo {year} {2012})}\BibitemShut {NoStop}%
\bibitem [{\citenamefont {Li}\ \emph {et~al.}(2003)\citenamefont {Li},
  \citenamefont {Wu}, \citenamefont {Kim}, \citenamefont {Shi}, \citenamefont
  {Yang},\ and\ \citenamefont {Majumdar}}]{ref:li2003tci}%
  \BibitemOpen
  \bibfield  {author} {\bibinfo {author} {\bibfnamefont {D.}~\bibnamefont
  {Li}}, \bibinfo {author} {\bibfnamefont {Y.}~\bibnamefont {Wu}}, \bibinfo
  {author} {\bibfnamefont {P.}~\bibnamefont {Kim}}, \bibinfo {author}
  {\bibfnamefont {L.}~\bibnamefont {Shi}}, \bibinfo {author} {\bibfnamefont
  {P.}~\bibnamefont {Yang}}, \ and\ \bibinfo {author} {\bibfnamefont
  {A.}~\bibnamefont {Majumdar}},\ }\href {\doibase
  http://dx.doi.org/10.1063/1.1616981} {\bibfield  {journal} {\bibinfo
  {journal} {Appl. Phys. Lett.}\ }\textbf {\bibinfo {volume} {83}},\ \bibinfo
  {pages} {2934} (\bibinfo {year} {2003})}\BibitemShut {NoStop}%
\bibitem [{\citenamefont {Hausmann}\ \emph {et~al.}(2014)\citenamefont
  {Hausmann}, \citenamefont {Bulu}, \citenamefont {Venkataraman}, \citenamefont
  {Deotare},\ and\ \citenamefont {Lon\v{c}ar}}]{ref:hausmann2014dnp}%
  \BibitemOpen
  \bibfield  {author} {\bibinfo {author} {\bibfnamefont {B.~J.}\ \bibnamefont
  {Hausmann}}, \bibinfo {author} {\bibfnamefont {I.}~\bibnamefont {Bulu}},
  \bibinfo {author} {\bibfnamefont {V.}~\bibnamefont {Venkataraman}}, \bibinfo
  {author} {\bibfnamefont {P.}~\bibnamefont {Deotare}}, \ and\ \bibinfo
  {author} {\bibfnamefont {M.}~\bibnamefont {Lon\v{c}ar}},\ }\href@noop {}
  {\bibfield  {journal} {\bibinfo  {journal} {Nature Photon.}\ }\textbf
  {\bibinfo {volume} {8}},\ \bibinfo {pages} {369} (\bibinfo {year}
  {2014})}\BibitemShut {NoStop}%
\bibitem [{\citenamefont {Chan}\ \emph {et~al.}(2012)\citenamefont {Chan},
  \citenamefont {Safavi-Naeini}, \citenamefont {Hill}, \citenamefont
  {Meenehan},\ and\ \citenamefont {Painter}}]{ref:chan2012ooc}%
  \BibitemOpen
  \bibfield  {author} {\bibinfo {author} {\bibfnamefont {J.}~\bibnamefont
  {Chan}}, \bibinfo {author} {\bibfnamefont {A.}~\bibnamefont {Safavi-Naeini}},
  \bibinfo {author} {\bibfnamefont {J.}~\bibnamefont {Hill}}, \bibinfo {author}
  {\bibfnamefont {S.}~\bibnamefont {Meenehan}}, \ and\ \bibinfo {author}
  {\bibfnamefont {O.}~\bibnamefont {Painter}},\ }\href@noop {} {\bibfield
  {journal} {\bibinfo  {journal} {Appl. Phys. Lett.}\ }\textbf {\bibinfo
  {volume} {101}},\ \bibinfo {pages} {081115} (\bibinfo {year}
  {2012})}\BibitemShut {NoStop}%
\bibitem [{\citenamefont {Hounsome}\ \emph {et~al.}(2006)\citenamefont
  {Hounsome}, \citenamefont {Jones}, \citenamefont {Shaw},\ and\ \citenamefont
  {Briddon}}]{ref:hounsome2006pcd}%
  \BibitemOpen
  \bibfield  {author} {\bibinfo {author} {\bibfnamefont {L.~S.}\ \bibnamefont
  {Hounsome}}, \bibinfo {author} {\bibfnamefont {R.}~\bibnamefont {Jones}},
  \bibinfo {author} {\bibfnamefont {M.~J.}\ \bibnamefont {Shaw}}, \ and\
  \bibinfo {author} {\bibfnamefont {P.~R.}\ \bibnamefont {Briddon}},\ }\href
  {\doibase 10.1002/pssa.200671121} {\bibfield  {journal} {\bibinfo  {journal}
  {Phys. Status Solidi A}\ }\textbf {\bibinfo {volume} {203}},\ \bibinfo
  {pages} {3088} (\bibinfo {year} {2006})}\BibitemShut {NoStop}%
\bibitem [{\citenamefont {Nguyen}\ \emph {et~al.}(2015)\citenamefont {Nguyen},
  \citenamefont {Hease}, \citenamefont {Baker}, \citenamefont {Gil-Santos},
  \citenamefont {Senellart}, \citenamefont {Lemaître}, \citenamefont {Ducci},
  \citenamefont {Leo},\ and\ \citenamefont {Favero}}]{ref:nguyen2015iod}%
  \BibitemOpen
  \bibfield  {author} {\bibinfo {author} {\bibfnamefont {D.~T.}\ \bibnamefont
  {Nguyen}}, \bibinfo {author} {\bibfnamefont {W.}~\bibnamefont {Hease}},
  \bibinfo {author} {\bibfnamefont {C.}~\bibnamefont {Baker}}, \bibinfo
  {author} {\bibfnamefont {E.}~\bibnamefont {Gil-Santos}}, \bibinfo {author}
  {\bibfnamefont {P.}~\bibnamefont {Senellart}}, \bibinfo {author}
  {\bibfnamefont {A.}~\bibnamefont {Lemaître}}, \bibinfo {author}
  {\bibfnamefont {S.}~\bibnamefont {Ducci}}, \bibinfo {author} {\bibfnamefont
  {G.}~\bibnamefont {Leo}}, \ and\ \bibinfo {author} {\bibfnamefont
  {I.}~\bibnamefont {Favero}},\ }\href@noop {} {\bibfield  {journal} {\bibinfo
  {journal} {New J. Phys.}\ }\textbf {\bibinfo {volume} {17}},\ \bibinfo
  {pages} {023016} (\bibinfo {year} {2015})}\BibitemShut {NoStop}%
\bibitem [{\citenamefont {Clark}\ \emph {et~al.}(2005)\citenamefont {Clark},
  \citenamefont {Hsu}, \citenamefont {Abdelmoneum},\ and\ \citenamefont
  {Nguyen}}]{ref:clark2005hqu}%
  \BibitemOpen
  \bibfield  {author} {\bibinfo {author} {\bibfnamefont {J.~R.}\ \bibnamefont
  {Clark}}, \bibinfo {author} {\bibfnamefont {W.-T.}\ \bibnamefont {Hsu}},
  \bibinfo {author} {\bibfnamefont {M.~A.}\ \bibnamefont {Abdelmoneum}}, \ and\
  \bibinfo {author} {\bibfnamefont {C.~T.-C.}\ \bibnamefont {Nguyen}},\
  }\href@noop {} {\bibfield  {journal} {\bibinfo  {journal} {Jour. of Micro.
  Sys.}\ }\textbf {\bibinfo {volume} {14}},\ \bibinfo {pages} {1298} (\bibinfo
  {year} {2005})}\BibitemShut {NoStop}%
\bibitem [{\citenamefont {Yang}\ \emph {et~al.}(2014)\citenamefont {Yang},
  \citenamefont {Wang}, \citenamefont {Lee}, \citenamefont {Ladhane},
  \citenamefont {Young},\ and\ \citenamefont {Feng}}]{ref:yang2014tdo}%
  \BibitemOpen
  \bibfield  {author} {\bibinfo {author} {\bibfnamefont {R.}~\bibnamefont
  {Yang}}, \bibinfo {author} {\bibfnamefont {Z.}~\bibnamefont {Wang}}, \bibinfo
  {author} {\bibfnamefont {J.}~\bibnamefont {Lee}}, \bibinfo {author}
  {\bibfnamefont {K.}~\bibnamefont {Ladhane}}, \bibinfo {author} {\bibfnamefont
  {D.~J.}\ \bibnamefont {Young}}, \ and\ \bibinfo {author} {\bibfnamefont
  {P.~X.~L.}\ \bibnamefont {Feng}},\ }in\ \href@noop {} {\emph {\bibinfo
  {booktitle} {Frequency Control Symposium (FCS), 2014 IEEE International}}}\
  (\bibinfo {year} {2014})\ pp.\ \bibinfo {pages} {1--3}\BibitemShut {NoStop}%
\bibitem [{\citenamefont {Jiang}\ \emph {et~al.}(2012)\citenamefont {Jiang},
  \citenamefont {Lu}, \citenamefont {Zhang},\ and\ \citenamefont
  {Lin}}]{ref:wei2012hfs}%
  \BibitemOpen
  \bibfield  {author} {\bibinfo {author} {\bibfnamefont {W.~C.}\ \bibnamefont
  {Jiang}}, \bibinfo {author} {\bibfnamefont {X.}~\bibnamefont {Lu}}, \bibinfo
  {author} {\bibfnamefont {J.}~\bibnamefont {Zhang}}, \ and\ \bibinfo {author}
  {\bibfnamefont {Q.}~\bibnamefont {Lin}},\ }\href {\doibase
  10.1364/OE.20.015991} {\bibfield  {journal} {\bibinfo  {journal} {Opt.
  Express}\ }\textbf {\bibinfo {volume} {20}},\ \bibinfo {pages} {15991}
  (\bibinfo {year} {2012})}\BibitemShut {NoStop}%
\bibitem [{\citenamefont {Hodges}\ \emph {et~al.}(2012)\citenamefont {Hodges},
  \citenamefont {Li}, \citenamefont {Lu}, \citenamefont {Chen}, \citenamefont
  {Trusheim}, \citenamefont {Allegri}, \citenamefont {Yao}, \citenamefont
  {Gaathon.}, \citenamefont {Bakhru},\ and\ \citenamefont
  {Englund}}]{ref:hodges2012lln}%
  \BibitemOpen
  \bibfield  {author} {\bibinfo {author} {\bibfnamefont {J.}~\bibnamefont
  {Hodges}}, \bibinfo {author} {\bibfnamefont {L.}~\bibnamefont {Li}}, \bibinfo
  {author} {\bibfnamefont {M.}~\bibnamefont {Lu}}, \bibinfo {author}
  {\bibfnamefont {E.~H.}\ \bibnamefont {Chen}}, \bibinfo {author}
  {\bibfnamefont {M.}~\bibnamefont {Trusheim}}, \bibinfo {author}
  {\bibfnamefont {S.}~\bibnamefont {Allegri}}, \bibinfo {author} {\bibfnamefont
  {X.}~\bibnamefont {Yao}}, \bibinfo {author} {\bibfnamefont {O.}~\bibnamefont
  {Gaathon.}}, \bibinfo {author} {\bibfnamefont {H.}~\bibnamefont {Bakhru}}, \
  and\ \bibinfo {author} {\bibfnamefont {D.}~\bibnamefont {Englund}},\
  }\href@noop {} {\bibfield  {journal} {\bibinfo  {journal} {New J. Phys.}\
  }\textbf {\bibinfo {volume} {14}},\ \bibinfo {pages} {093004} (\bibinfo
  {year} {2012})}\BibitemShut {NoStop}%
\bibitem [{\citenamefont {Bar-Gill}\ \emph {et~al.}(2013)\citenamefont
  {Bar-Gill}, \citenamefont {Pham}, \citenamefont {Jarmola}, \citenamefont
  {Budker},\ and\ \citenamefont {Walsworth}}]{ref:bar-gill2013ses}%
  \BibitemOpen
  \bibfield  {author} {\bibinfo {author} {\bibfnamefont {N.}~\bibnamefont
  {Bar-Gill}}, \bibinfo {author} {\bibfnamefont {L.}~\bibnamefont {Pham}},
  \bibinfo {author} {\bibfnamefont {A.}~\bibnamefont {Jarmola}}, \bibinfo
  {author} {\bibfnamefont {D.}~\bibnamefont {Budker}}, \ and\ \bibinfo {author}
  {\bibfnamefont {R.}~\bibnamefont {Walsworth}},\ }\href@noop {} {\bibfield
  {journal} {\bibinfo  {journal} {Nat. Commun.}\ }\textbf {\bibinfo {volume}
  {4}},\ \bibinfo {pages} {1743} (\bibinfo {year} {2013})}\BibitemShut
  {NoStop}%
\bibitem [{\citenamefont {Davies}\ and\ \citenamefont
  {Hamer}(1976)}]{ref:davies1976oso}%
  \BibitemOpen
  \bibfield  {author} {\bibinfo {author} {\bibfnamefont {G.}~\bibnamefont
  {Davies}}\ and\ \bibinfo {author} {\bibfnamefont {M.~F.}\ \bibnamefont
  {Hamer}},\ }\href@noop {} {\bibfield  {journal} {\bibinfo  {journal} {Proc.
  R. Soc. Lond.~A}\ }\textbf {\bibinfo {volume} {348}},\ \bibinfo {pages} {285}
  (\bibinfo {year} {1976})}\BibitemShut {NoStop}%
\bibitem [{\citenamefont {Batalov}\ \emph {et~al.}(2009)\citenamefont
  {Batalov}, \citenamefont {Jacques}, \citenamefont {Kaiser}, \citenamefont
  {Siyushev}, \citenamefont {Neumann}, \citenamefont {Rogers}, \citenamefont
  {McMurtrie}, \citenamefont {Manson}, \citenamefont {Jelezko},\ and\
  \citenamefont {Wrachtrup}}]{ref:batalov2009lts}%
  \BibitemOpen
  \bibfield  {author} {\bibinfo {author} {\bibfnamefont {A.}~\bibnamefont
  {Batalov}}, \bibinfo {author} {\bibfnamefont {V.}~\bibnamefont {Jacques}},
  \bibinfo {author} {\bibfnamefont {F.}~\bibnamefont {Kaiser}}, \bibinfo
  {author} {\bibfnamefont {P.}~\bibnamefont {Siyushev}}, \bibinfo {author}
  {\bibfnamefont {P.}~\bibnamefont {Neumann}}, \bibinfo {author} {\bibfnamefont
  {L.~J.}\ \bibnamefont {Rogers}}, \bibinfo {author} {\bibfnamefont {R.~L.}\
  \bibnamefont {McMurtrie}}, \bibinfo {author} {\bibfnamefont {N.~B.}\
  \bibnamefont {Manson}}, \bibinfo {author} {\bibfnamefont {F.}~\bibnamefont
  {Jelezko}}, \ and\ \bibinfo {author} {\bibfnamefont {J.}~\bibnamefont
  {Wrachtrup}},\ }\href {\doibase 10.1103/PhysRevLett.102.195506} {\bibfield
  {journal} {\bibinfo  {journal} {Phys. Rev. Lett.}\ }\textbf {\bibinfo
  {volume} {102}},\ \bibinfo {pages} {195506} (\bibinfo {year}
  {2009})}\BibitemShut {NoStop}%
\bibitem [{\citenamefont {Baker}\ \emph {et~al.}(2014)\citenamefont {Baker},
  \citenamefont {Hease}, \citenamefont {Nguyen}, \citenamefont {Andronico},
  \citenamefont {Ducci}, \citenamefont {Leo},\ and\ \citenamefont
  {Favero}}]{ref:baker2014pcg}%
  \BibitemOpen
  \bibfield  {author} {\bibinfo {author} {\bibfnamefont {C.}~\bibnamefont
  {Baker}}, \bibinfo {author} {\bibfnamefont {W.}~\bibnamefont {Hease}},
  \bibinfo {author} {\bibfnamefont {D.-T.}\ \bibnamefont {Nguyen}}, \bibinfo
  {author} {\bibfnamefont {A.}~\bibnamefont {Andronico}}, \bibinfo {author}
  {\bibfnamefont {S.}~\bibnamefont {Ducci}}, \bibinfo {author} {\bibfnamefont
  {G.}~\bibnamefont {Leo}}, \ and\ \bibinfo {author} {\bibfnamefont
  {I.}~\bibnamefont {Favero}},\ }\href {\doibase 10.1364/OE.22.014072}
  {\bibfield  {journal} {\bibinfo  {journal} {Opt. Express}\ }\textbf {\bibinfo
  {volume} {22}},\ \bibinfo {pages} {14072} (\bibinfo {year}
  {2014})}\BibitemShut {NoStop}%
\bibitem [{\citenamefont {Dong}\ \emph {et~al.}(2012)\citenamefont {Dong},
  \citenamefont {Fiore}, \citenamefont {Kuzyk},\ and\ \citenamefont
  {Wang}}]{ref:dong2012odm}%
  \BibitemOpen
  \bibfield  {author} {\bibinfo {author} {\bibfnamefont {C.}~\bibnamefont
  {Dong}}, \bibinfo {author} {\bibfnamefont {V.}~\bibnamefont {Fiore}},
  \bibinfo {author} {\bibfnamefont {M.~C.}\ \bibnamefont {Kuzyk}}, \ and\
  \bibinfo {author} {\bibfnamefont {H.}~\bibnamefont {Wang}},\ }\href@noop {}
  {\bibfield  {journal} {\bibinfo  {journal} {Science}\ }\textbf {\bibinfo
  {volume} {338}},\ \bibinfo {pages} {1609} (\bibinfo {year}
  {2012})}\BibitemShut {NoStop}%
\bibitem [{\citenamefont {Hill}\ \emph {et~al.}(2012)\citenamefont {Hill},
  \citenamefont {Safavi-Naeini}, \citenamefont {Chan},\ and\ \citenamefont
  {Painter}}]{ref:hill2012cow}%
  \BibitemOpen
  \bibfield  {author} {\bibinfo {author} {\bibfnamefont {J.~T.}\ \bibnamefont
  {Hill}}, \bibinfo {author} {\bibfnamefont {A.~H.}\ \bibnamefont
  {Safavi-Naeini}}, \bibinfo {author} {\bibfnamefont {J.}~\bibnamefont {Chan}},
  \ and\ \bibinfo {author} {\bibfnamefont {O.}~\bibnamefont {Painter}},\
  }\href@noop {} {\bibfield  {journal} {\bibinfo  {journal} {Nat. Commun.}\
  }\textbf {\bibinfo {volume} {3}},\ \bibinfo {pages} {1196} (\bibinfo {year}
  {2012})}\BibitemShut {NoStop}%
\bibitem [{\citenamefont {Stannigel}\ \emph {et~al.}(2012)\citenamefont
  {Stannigel}, \citenamefont {Komar}, \citenamefont {Habraken}, \citenamefont
  {Bennett}, \citenamefont {Lukin}, \citenamefont {Zoller},\ and\ \citenamefont
  {Rabl}}]{ref:stannigel2012oqi}%
  \BibitemOpen
  \bibfield  {author} {\bibinfo {author} {\bibfnamefont {K.}~\bibnamefont
  {Stannigel}}, \bibinfo {author} {\bibfnamefont {P.}~\bibnamefont {Komar}},
  \bibinfo {author} {\bibfnamefont {S.~J.~M.}\ \bibnamefont {Habraken}},
  \bibinfo {author} {\bibfnamefont {S.~D.}\ \bibnamefont {Bennett}}, \bibinfo
  {author} {\bibfnamefont {M.~D.}\ \bibnamefont {Lukin}}, \bibinfo {author}
  {\bibfnamefont {P.}~\bibnamefont {Zoller}}, \ and\ \bibinfo {author}
  {\bibfnamefont {P.}~\bibnamefont {Rabl}},\ }\href {\doibase
  10.1103/PhysRevLett.109.013603} {\bibfield  {journal} {\bibinfo  {journal}
  {Phys. Rev. Lett.}\ }\textbf {\bibinfo {volume} {109}},\ \bibinfo {pages}
  {013603} (\bibinfo {year} {2012})}\BibitemShut {NoStop}%
\bibitem [{\citenamefont {Burek}\ \emph {et~al.}(2015)\citenamefont {Burek},
  \citenamefont {Cohen}, \citenamefont {Meenehan}, \citenamefont {Ruelle},
  \citenamefont {Meesala}, \citenamefont {Rochman}, \citenamefont {Atikian},
  \citenamefont {Markham}, \citenamefont {Twitchen}, \citenamefont {Lukin},
  \citenamefont {Painter},\ and\ \citenamefont
  {Lon\v{c}ar}}]{ref:burek2015doc}%
  \BibitemOpen
  \bibfield  {author} {\bibinfo {author} {\bibfnamefont {M.~J.}\ \bibnamefont
  {Burek}}, \bibinfo {author} {\bibfnamefont {J.~D.}\ \bibnamefont {Cohen}},
  \bibinfo {author} {\bibfnamefont {S.~M.}\ \bibnamefont {Meenehan}}, \bibinfo
  {author} {\bibfnamefont {T.}~\bibnamefont {Ruelle}}, \bibinfo {author}
  {\bibfnamefont {S.}~\bibnamefont {Meesala}}, \bibinfo {author} {\bibfnamefont
  {J.}~\bibnamefont {Rochman}}, \bibinfo {author} {\bibfnamefont {H.~A.}\
  \bibnamefont {Atikian}}, \bibinfo {author} {\bibfnamefont {M.}~\bibnamefont
  {Markham}}, \bibinfo {author} {\bibfnamefont {D.~J.}\ \bibnamefont
  {Twitchen}}, \bibinfo {author} {\bibfnamefont {M.~D.}\ \bibnamefont {Lukin}},
  \bibinfo {author} {\bibfnamefont {O.}~\bibnamefont {Painter}}, \ and\
  \bibinfo {author} {\bibfnamefont {M.}~\bibnamefont {Lon\v{c}ar}},\
  }\href@noop {} {\bibfield  {journal} {\bibinfo  {journal} {arXiv:1512.04166}\
  } (\bibinfo {year} {2015})}\BibitemShut {NoStop}%
\end{thebibliography}

\begin{thebibliography}{24}%
\makeatletter
\providecommand \@ifxundefined [1]{%
 \@ifx{#1\undefined}
}%
\providecommand \@ifnum [1]{%
 \ifnum #1\expandafter \@firstoftwo
 \else \expandafter \@secondoftwo
 \fi
}%
\providecommand \@ifx [1]{%
 \ifx #1\expandafter \@firstoftwo
 \else \expandafter \@secondoftwo
 \fi
}%
\providecommand \natexlab [1]{#1}%
\providecommand \enquote  [1]{``#1''}%
\providecommand \bibnamefont  [1]{#1}%
\providecommand \bibfnamefont [1]{#1}%
\providecommand \citenamefont [1]{#1}%
\providecommand \href@noop [0]{\@secondoftwo}%
\providecommand \href [0]{\begingroup \@sanitize@url \@href}%
\providecommand \@href[1]{\@@startlink{#1}\@@href}%
\providecommand \@@href[1]{\endgroup#1\@@endlink}%
\providecommand \@sanitize@url [0]{\catcode `\\12\catcode `\$12\catcode
  `\&12\catcode `\#12\catcode `\^12\catcode `\_12\catcode `\%12\relax}%
\providecommand \@@startlink[1]{}%
\providecommand \@@endlink[0]{}%
\providecommand \url  [0]{\begingroup\@sanitize@url \@url }%
\providecommand \@url [1]{\endgroup\@href {#1}{\urlprefix }}%
\providecommand \urlprefix  [0]{URL }%
\providecommand \Eprint [0]{\href }%
\providecommand \doibase [0]{http://dx.doi.org/}%
\providecommand \selectlanguage [0]{\@gobble}%
\providecommand \bibinfo  [0]{\@secondoftwo}%
\providecommand \bibfield  [0]{\@secondoftwo}%
\providecommand \translation [1]{[#1]}%
\providecommand \BibitemOpen [0]{}%
\providecommand \bibitemStop [0]{}%
\providecommand \bibitemNoStop [0]{.\EOS\space}%
\providecommand \EOS [0]{\spacefactor3000\relax}%
\providecommand \BibitemShut  [1]{\csname bibitem#1\endcsname}%
\let\auto@bib@innerbib\@empty
\bibitem [{\citenamefont {Carmon}\ \emph {et~al.}(2004)\citenamefont {Carmon},
  \citenamefont {Yang},\ and\ \citenamefont {Vahala}}]{ref:supp_carmon2004dtb}%
  \BibitemOpen
  \bibfield  {author} {\bibinfo {author} {\bibfnamefont {T.}~\bibnamefont
  {Carmon}}, \bibinfo {author} {\bibfnamefont {L.}~\bibnamefont {Yang}}, \ and\
  \bibinfo {author} {\bibfnamefont {K.~J.}\ \bibnamefont {Vahala}},\ }\href
  {http://www.opticsexpress.org/abstract.cfm?URI=OPEX-12-20-4742} {\bibfield
  {journal} {\bibinfo  {journal} {Opt. Express}\ }\textbf {\bibinfo {volume}
  {12}},\ \bibinfo {pages} {4742} (\bibinfo {year} {2004})}\BibitemShut
  {NoStop}%
\bibitem [{\citenamefont {Li}\ \emph {et~al.}(2012)\citenamefont {Li},
  \citenamefont {Mingo}, \citenamefont {Lindsay}, \citenamefont {Broido},
  \citenamefont {Stewart},\ and\ \citenamefont {Katcho}}]{ref:supp_li2012tcd}%
  \BibitemOpen
  \bibfield  {author} {\bibinfo {author} {\bibfnamefont {W.}~\bibnamefont
  {Li}}, \bibinfo {author} {\bibfnamefont {N.}~\bibnamefont {Mingo}}, \bibinfo
  {author} {\bibfnamefont {L.}~\bibnamefont {Lindsay}}, \bibinfo {author}
  {\bibfnamefont {D.~A.}\ \bibnamefont {Broido}}, \bibinfo {author}
  {\bibfnamefont {D.~A.}\ \bibnamefont {Stewart}}, \ and\ \bibinfo {author}
  {\bibfnamefont {N.~A.}\ \bibnamefont {Katcho}},\ }\href {\doibase
  10.1103/PhysRevB.85.195436} {\bibfield  {journal} {\bibinfo  {journal} {Phys.
  Rev. B}\ }\textbf {\bibinfo {volume} {85}},\ \bibinfo {pages} {195436}
  (\bibinfo {year} {2012})}\BibitemShut {NoStop}%
\bibitem [{\citenamefont {Aspelmeyer}\ \emph {et~al.}(2014)\citenamefont
  {Aspelmeyer}, \citenamefont {Kippenberg},\ and\ \citenamefont
  {Marquardt}}]{ref:supp_aspelmeyer2014co}%
  \BibitemOpen
  \bibfield  {author} {\bibinfo {author} {\bibfnamefont {M.}~\bibnamefont
  {Aspelmeyer}}, \bibinfo {author} {\bibfnamefont {T.~J.}\ \bibnamefont
  {Kippenberg}}, \ and\ \bibinfo {author} {\bibfnamefont {F.}~\bibnamefont
  {Marquardt}},\ }\href {\doibase 10.1103/RevModPhys.86.1391} {\bibfield
  {journal} {\bibinfo  {journal} {Rev. Mod. Phys.}\ }\textbf {\bibinfo {volume}
  {86}},\ \bibinfo {pages} {1391} (\bibinfo {year} {2014})}\BibitemShut
  {NoStop}%
\bibitem [{\citenamefont {Lin}\ \emph {et~al.}(2010)\citenamefont {Lin},
  \citenamefont {Rosenberg}, \citenamefont {Chang}, \citenamefont {Camacho},
  \citenamefont {Eichenfield}, \citenamefont {Vahala},\ and\ \citenamefont
  {Painter}}]{ref:supp_lin2010cmo}%
  \BibitemOpen
  \bibfield  {author} {\bibinfo {author} {\bibfnamefont {Q.}~\bibnamefont
  {Lin}}, \bibinfo {author} {\bibfnamefont {J.}~\bibnamefont {Rosenberg}},
  \bibinfo {author} {\bibfnamefont {D.}~\bibnamefont {Chang}}, \bibinfo
  {author} {\bibfnamefont {R.}~\bibnamefont {Camacho}}, \bibinfo {author}
  {\bibfnamefont {M.}~\bibnamefont {Eichenfield}}, \bibinfo {author}
  {\bibfnamefont {K.~J.}\ \bibnamefont {Vahala}}, \ and\ \bibinfo {author}
  {\bibfnamefont {O.}~\bibnamefont {Painter}},\ }\href {\doibase
  10.1038/nphoton.2010.5} {\bibfield  {journal} {\bibinfo  {journal} {Nature
  Photon.}\ }\textbf {\bibinfo {volume} {4}},\ \bibinfo {pages} {236} (\bibinfo
  {year} {2010})}\BibitemShut {NoStop}%
\bibitem [{\citenamefont {Lu}\ \emph {et~al.}(2015)\citenamefont {Lu},
  \citenamefont {Lee},\ and\ \citenamefont {Lin}}]{ref:supp_lu2015hfh}%
  \BibitemOpen
  \bibfield  {author} {\bibinfo {author} {\bibfnamefont {X.}~\bibnamefont
  {Lu}}, \bibinfo {author} {\bibfnamefont {J.~Y.}\ \bibnamefont {Lee}}, \ and\
  \bibinfo {author} {\bibfnamefont {Q.}~\bibnamefont {Lin}},\ }\href@noop {}
  {\bibfield  {journal} {\bibinfo  {journal} {Sci. Rep.}\ }\textbf {\bibinfo
  {volume} {5}},\ \bibinfo {pages} {17005} (\bibinfo {year}
  {2015})}\BibitemShut {NoStop}%
\bibitem [{\citenamefont {Nguyen}\ \emph {et~al.}(2013)\citenamefont {Nguyen},
  \citenamefont {Baker}, \citenamefont {Hease}, \citenamefont {Sejil},
  \citenamefont {Senellart}, \citenamefont {Lema\^{i}tre}, \citenamefont
  {Ducci}, \citenamefont {Leo},\ and\ \citenamefont
  {Favero}}]{ref:supp_nguyen2013uqf}%
  \BibitemOpen
  \bibfield  {author} {\bibinfo {author} {\bibfnamefont {D.~T.}\ \bibnamefont
  {Nguyen}}, \bibinfo {author} {\bibfnamefont {C.}~\bibnamefont {Baker}},
  \bibinfo {author} {\bibfnamefont {W.}~\bibnamefont {Hease}}, \bibinfo
  {author} {\bibfnamefont {S.}~\bibnamefont {Sejil}}, \bibinfo {author}
  {\bibfnamefont {P.}~\bibnamefont {Senellart}}, \bibinfo {author}
  {\bibfnamefont {A.}~\bibnamefont {Lema\^{i}tre}}, \bibinfo {author}
  {\bibfnamefont {S.}~\bibnamefont {Ducci}}, \bibinfo {author} {\bibfnamefont
  {G.}~\bibnamefont {Leo}}, \ and\ \bibinfo {author} {\bibfnamefont
  {I.}~\bibnamefont {Favero}},\ }\href@noop {} {\bibfield  {journal} {\bibinfo
  {journal} {Appl. Phys. Lett.}\ }\textbf {\bibinfo {volume} {103}},\ \bibinfo
  {eid} {241112} (\bibinfo {year} {2013})}\BibitemShut {NoStop}%
\bibitem [{\citenamefont {Mitchell}\ \emph {et~al.}(2014)\citenamefont
  {Mitchell}, \citenamefont {Hryciw},\ and\ \citenamefont
  {Barclay}}]{ref:supp_mitchell2014cog}%
  \BibitemOpen
  \bibfield  {author} {\bibinfo {author} {\bibfnamefont {M.}~\bibnamefont
  {Mitchell}}, \bibinfo {author} {\bibfnamefont {A.~C.}\ \bibnamefont
  {Hryciw}}, \ and\ \bibinfo {author} {\bibfnamefont {P.~E.}\ \bibnamefont
  {Barclay}},\ }\href
  {http://scitation.aip.org/content/aip/journal/apl/104/14/10.1063/1.4870999}
  {\bibfield  {journal} {\bibinfo  {journal} {Appl. Phys. Lett.}\ }\textbf
  {\bibinfo {volume} {104}},\ \bibinfo {eid} {141104} (\bibinfo {year}
  {2014})}\BibitemShut {NoStop}%
\bibitem [{\citenamefont {Liu}\ \emph {et~al.}(2013)\citenamefont {Liu},
  \citenamefont {Davan{\c{c}}o}, \citenamefont {Aksyuk},\ and\ \citenamefont
  {Srinivasan}}]{ref:supp_liu2013eit}%
  \BibitemOpen
  \bibfield  {author} {\bibinfo {author} {\bibfnamefont {Y.}~\bibnamefont
  {Liu}}, \bibinfo {author} {\bibfnamefont {M.}~\bibnamefont {Davan{\c{c}}o}},
  \bibinfo {author} {\bibfnamefont {V.}~\bibnamefont {Aksyuk}}, \ and\ \bibinfo
  {author} {\bibfnamefont {K.}~\bibnamefont {Srinivasan}},\ }\href@noop {}
  {\bibfield  {journal} {\bibinfo  {journal} {Phys. Rev. Lett.}\ }\textbf
  {\bibinfo {volume} {110}},\ \bibinfo {pages} {223603} (\bibinfo {year}
  {2013})}\BibitemShut {NoStop}%
\bibitem [{\citenamefont {Fong}\ \emph {et~al.}(2012)\citenamefont {Fong},
  \citenamefont {Pernice},\ and\ \citenamefont {Tang}}]{ref:supp_fong2012fpn}%
  \BibitemOpen
  \bibfield  {author} {\bibinfo {author} {\bibfnamefont {K.~Y.}\ \bibnamefont
  {Fong}}, \bibinfo {author} {\bibfnamefont {W.~H.~P.}\ \bibnamefont
  {Pernice}}, \ and\ \bibinfo {author} {\bibfnamefont {H.~X.}\ \bibnamefont
  {Tang}},\ }\href {\doibase 10.1103/PhysRevB.85.161410} {\bibfield  {journal}
  {\bibinfo  {journal} {Phys. Rev. B}\ }\textbf {\bibinfo {volume} {85}},\
  \bibinfo {pages} {161410} (\bibinfo {year} {2012})}\BibitemShut {NoStop}%
\bibitem [{\citenamefont {Grutter}\ \emph {et~al.}(2015)\citenamefont
  {Grutter}, \citenamefont {Davan\c{c}o},\ and\ \citenamefont
  {Srinivasan}}]{ref:supp_grutter2015soc}%
  \BibitemOpen
  \bibfield  {author} {\bibinfo {author} {\bibfnamefont {K.~E.}\ \bibnamefont
  {Grutter}}, \bibinfo {author} {\bibfnamefont {M.~I.}\ \bibnamefont
  {Davan\c{c}o}}, \ and\ \bibinfo {author} {\bibfnamefont {K.}~\bibnamefont
  {Srinivasan}},\ }\href {\doibase 10.1364/OPTICA.2.000994} {\bibfield
  {journal} {\bibinfo  {journal} {Optica}\ }\textbf {\bibinfo {volume} {2}},\
  \bibinfo {pages} {994} (\bibinfo {year} {2015})}\BibitemShut {NoStop}%
\bibitem [{\citenamefont {Xiong}\ \emph {et~al.}(2012)\citenamefont {Xiong},
  \citenamefont {Sun}, \citenamefont {Fong},\ and\ \citenamefont
  {Tang}}]{ref:supp_xiong2012ihf}%
  \BibitemOpen
  \bibfield  {author} {\bibinfo {author} {\bibfnamefont {C.}~\bibnamefont
  {Xiong}}, \bibinfo {author} {\bibfnamefont {X.}~\bibnamefont {Sun}}, \bibinfo
  {author} {\bibfnamefont {K.~Y.}\ \bibnamefont {Fong}}, \ and\ \bibinfo
  {author} {\bibfnamefont {H.~X.}\ \bibnamefont {Tang}},\ }\href@noop {}
  {\bibfield  {journal} {\bibinfo  {journal} {Appl. Phys. Lett.}\ }\textbf
  {\bibinfo {volume} {100}},\ \bibinfo {pages} {171111} (\bibinfo {year}
  {2012})}\BibitemShut {NoStop}%
\bibitem [{\citenamefont {Bochmann}\ \emph {et~al.}(2013)\citenamefont
  {Bochmann}, \citenamefont {Vainsencher}, \citenamefont {Awschalom},\ and\
  \citenamefont {Cleland}}]{ref:supp_bochmann2013ncb}%
  \BibitemOpen
  \bibfield  {author} {\bibinfo {author} {\bibfnamefont {J.}~\bibnamefont
  {Bochmann}}, \bibinfo {author} {\bibfnamefont {A.}~\bibnamefont
  {Vainsencher}}, \bibinfo {author} {\bibfnamefont {D.~D.}\ \bibnamefont
  {Awschalom}}, \ and\ \bibinfo {author} {\bibfnamefont {A.~N.}\ \bibnamefont
  {Cleland}},\ }\href@noop {} {\bibfield  {journal} {\bibinfo  {journal}
  {Nature Phys.}\ }\textbf {\bibinfo {volume} {9}},\ \bibinfo {pages}
  {712–716} (\bibinfo {year} {2013})}\BibitemShut {NoStop}%
\bibitem [{\citenamefont {Eichenfield}\ \emph {et~al.}(2009)\citenamefont
  {Eichenfield}, \citenamefont {Chan}, \citenamefont {Camacho}, \citenamefont
  {Vahala},\ and\ \citenamefont {Painter}}]{ref:supp_eichenfield2009oc}%
  \BibitemOpen
  \bibfield  {author} {\bibinfo {author} {\bibfnamefont {M.}~\bibnamefont
  {Eichenfield}}, \bibinfo {author} {\bibfnamefont {J.}~\bibnamefont {Chan}},
  \bibinfo {author} {\bibfnamefont {R.}~\bibnamefont {Camacho}}, \bibinfo
  {author} {\bibfnamefont {K.}~\bibnamefont {Vahala}}, \ and\ \bibinfo {author}
  {\bibfnamefont {O.}~\bibnamefont {Painter}},\ }\href@noop {} {\bibfield
  {journal} {\bibinfo  {journal} {Nature}\ }\textbf {\bibinfo {volume} {462}},\
  \bibinfo {pages} {78} (\bibinfo {year} {2009})}\BibitemShut {NoStop}%
\bibitem [{\citenamefont {Bui}\ \emph {et~al.}(2012)\citenamefont {Bui},
  \citenamefont {Zheng}, \citenamefont {Hoch}, \citenamefont {Lee},
  \citenamefont {Harris},\ and\ \citenamefont {Wei~Wong}}]{ref:supp_bui2012hrh}%
  \BibitemOpen
  \bibfield  {author} {\bibinfo {author} {\bibfnamefont {C.~H.}\ \bibnamefont
  {Bui}}, \bibinfo {author} {\bibfnamefont {J.}~\bibnamefont {Zheng}}, \bibinfo
  {author} {\bibfnamefont {S.~W.}\ \bibnamefont {Hoch}}, \bibinfo {author}
  {\bibfnamefont {L.~Y.~T.}\ \bibnamefont {Lee}}, \bibinfo {author}
  {\bibfnamefont {J.~G.~E.}\ \bibnamefont {Harris}}, \ and\ \bibinfo {author}
  {\bibfnamefont {C.}~\bibnamefont {Wei~Wong}},\ }\href@noop {} {\bibfield
  {journal} {\bibinfo  {journal} {Appl. Phys. Lett.}\ }\textbf {\bibinfo
  {volume} {100}},\ \bibinfo {eid} {021110} (\bibinfo {year}
  {2012})}\BibitemShut {NoStop}%
\bibitem [{\citenamefont {Wilson}\ \emph {et~al.}(2009)\citenamefont {Wilson},
  \citenamefont {Regal}, \citenamefont {Papp},\ and\ \citenamefont
  {Kimble}}]{wilson2009cos}%
  \BibitemOpen
  \bibfield  {author} {\bibinfo {author} {\bibfnamefont {D.~J.}\ \bibnamefont
  {Wilson}}, \bibinfo {author} {\bibfnamefont {C.~A.}\ \bibnamefont {Regal}},
  \bibinfo {author} {\bibfnamefont {S.~B.}\ \bibnamefont {Papp}}, \ and\
  \bibinfo {author} {\bibfnamefont {H.~J.}\ \bibnamefont {Kimble}},\ }\href
  {\doibase 10.1103/PhysRevLett.103.207204} {\bibfield  {journal} {\bibinfo
  {journal} {Phys. Rev. Lett.}\ }\textbf {\bibinfo {volume} {103}},\ \bibinfo
  {pages} {207204} (\bibinfo {year} {2009})}\BibitemShut {NoStop}%
\bibitem [{\citenamefont {Reinhardt}\ \emph {et~al.}(2016)\citenamefont
  {Reinhardt}, \citenamefont {M\"{u}ller}, \citenamefont {Bourassa},\ and\
  \citenamefont {Sankey}}]{ref:supp_reinhardt2016uns}%
  \BibitemOpen
  \bibfield  {author} {\bibinfo {author} {\bibfnamefont {C.}~\bibnamefont
  {Reinhardt}}, \bibinfo {author} {\bibfnamefont {T.}~\bibnamefont
  {M\"{u}ller}}, \bibinfo {author} {\bibfnamefont {A.}~\bibnamefont
  {Bourassa}}, \ and\ \bibinfo {author} {\bibfnamefont {J.~C.}\ \bibnamefont
  {Sankey}},\ }\href@noop {} {\bibfield  {journal} {\bibinfo  {journal}
  {arXiv:1511.01769}\ } (\bibinfo {year} {2016})}\BibitemShut {NoStop}%
\bibitem [{\citenamefont {Norte}\ \emph {et~al.}(2016)\citenamefont {Norte},
  \citenamefont {Moura},\ and\ \citenamefont
  {Gr\"oblacher}}]{ref:supp_norte2016mrq}%
  \BibitemOpen
  \bibfield  {author} {\bibinfo {author} {\bibfnamefont {R.~A.}\ \bibnamefont
  {Norte}}, \bibinfo {author} {\bibfnamefont {J.~P.}\ \bibnamefont {Moura}}, \
  and\ \bibinfo {author} {\bibfnamefont {S.}~\bibnamefont {Gr\"oblacher}},\
  }\href {\doibase 10.1103/PhysRevLett.116.147202} {\bibfield  {journal}
  {\bibinfo  {journal} {Phys. Rev. Lett.}\ }\textbf {\bibinfo {volume} {116}},\
  \bibinfo {pages} {147202} (\bibinfo {year} {2016})}\BibitemShut {NoStop}%
\bibitem [{\citenamefont {Zhang}\ \emph {et~al.}(2015)\citenamefont {Zhang},
  \citenamefont {Ti}, \citenamefont {Davanço}, \citenamefont {Ren},
  \citenamefont {Aksyuk}, \citenamefont {Liu},\ and\ \citenamefont
  {Srinivasan}}]{ref:supp_zhang2015itf}%
  \BibitemOpen
  \bibfield  {author} {\bibinfo {author} {\bibfnamefont {R.}~\bibnamefont
  {Zhang}}, \bibinfo {author} {\bibfnamefont {C.}~\bibnamefont {Ti}}, \bibinfo
  {author} {\bibfnamefont {M.~I.}\ \bibnamefont {Davanço}}, \bibinfo {author}
  {\bibfnamefont {Y.}~\bibnamefont {Ren}}, \bibinfo {author} {\bibfnamefont
  {V.}~\bibnamefont {Aksyuk}}, \bibinfo {author} {\bibfnamefont
  {Y.}~\bibnamefont {Liu}}, \ and\ \bibinfo {author} {\bibfnamefont
  {K.}~\bibnamefont {Srinivasan}},\ }\href@noop {} {\bibfield  {journal}
  {\bibinfo  {journal} {Appl. Phys. Lett.}\ }\textbf {\bibinfo {volume}
  {107}},\ \bibinfo {eid} {131110} (\bibinfo {year} {2015})}\BibitemShut
  {NoStop}%
\bibitem [{\citenamefont {Chan}\ \emph {et~al.}(2011)\citenamefont {Chan},
  \citenamefont {Alegre}, \citenamefont {Safavi-Naeini}, \citenamefont {Hill},
  \citenamefont {Krause}, \citenamefont {Groblacher}, \citenamefont
  {Aspelmeyer},\ and\ \citenamefont {Painter}}]{ref:supp_chan2011lcn}%
  \BibitemOpen
  \bibfield  {author} {\bibinfo {author} {\bibfnamefont {J.}~\bibnamefont
  {Chan}}, \bibinfo {author} {\bibfnamefont {T.~P.~M.}\ \bibnamefont {Alegre}},
  \bibinfo {author} {\bibfnamefont {A.~H.}\ \bibnamefont {Safavi-Naeini}},
  \bibinfo {author} {\bibfnamefont {J.~T.}\ \bibnamefont {Hill}}, \bibinfo
  {author} {\bibfnamefont {A.}~\bibnamefont {Krause}}, \bibinfo {author}
  {\bibfnamefont {S.}~\bibnamefont {Groblacher}}, \bibinfo {author}
  {\bibfnamefont {M.}~\bibnamefont {Aspelmeyer}}, \ and\ \bibinfo {author}
  {\bibfnamefont {O.}~\bibnamefont {Painter}},\ }\href@noop {} {\bibfield
  {journal} {\bibinfo  {journal} {Nature}\ }\textbf {\bibinfo {volume} {478}},\
  \bibinfo {pages} {89} (\bibinfo {year} {2011})}\BibitemShut {NoStop}%
\bibitem [{\citenamefont {Krause}\ \emph {et~al.}(2015)\citenamefont {Krause},
  \citenamefont {Hill}, \citenamefont {Ludwig}, \citenamefont {Safavi-Naeini},
  \citenamefont {Chan}, \citenamefont {Marquardt},\ and\ \citenamefont
  {Painter}}]{ref:supp_krause2015nrp}%
  \BibitemOpen
  \bibfield  {author} {\bibinfo {author} {\bibfnamefont {A.~G.}\ \bibnamefont
  {Krause}}, \bibinfo {author} {\bibfnamefont {J.~T.}\ \bibnamefont {Hill}},
  \bibinfo {author} {\bibfnamefont {M.}~\bibnamefont {Ludwig}}, \bibinfo
  {author} {\bibfnamefont {A.~H.}\ \bibnamefont {Safavi-Naeini}}, \bibinfo
  {author} {\bibfnamefont {J.}~\bibnamefont {Chan}}, \bibinfo {author}
  {\bibfnamefont {F.}~\bibnamefont {Marquardt}}, \ and\ \bibinfo {author}
  {\bibfnamefont {O.}~\bibnamefont {Painter}},\ }\href@noop {} {\bibfield
  {journal} {\bibinfo  {journal} {Phys. Rev. Lett.}\ }\textbf {\bibinfo
  {volume} {115}},\ \bibinfo {pages} {233601} (\bibinfo {year}
  {2015})}\BibitemShut {NoStop}%
\bibitem [{\citenamefont {Meenehan}\ \emph {et~al.}(2015)\citenamefont
  {Meenehan}, \citenamefont {Cohen}, \citenamefont {MacCabe}, \citenamefont
  {Marsili}, \citenamefont {Shaw},\ and\ \citenamefont
  {Painter}}]{ref:supp_meenehan2015ped}%
  \BibitemOpen
  \bibfield  {author} {\bibinfo {author} {\bibfnamefont {S.~M.}\ \bibnamefont
  {Meenehan}}, \bibinfo {author} {\bibfnamefont {J.~D.}\ \bibnamefont {Cohen}},
  \bibinfo {author} {\bibfnamefont {G.~S.}\ \bibnamefont {MacCabe}}, \bibinfo
  {author} {\bibfnamefont {F.}~\bibnamefont {Marsili}}, \bibinfo {author}
  {\bibfnamefont {M.~D.}\ \bibnamefont {Shaw}}, \ and\ \bibinfo {author}
  {\bibfnamefont {O.}~\bibnamefont {Painter}},\ }\href@noop {} {\bibfield
  {journal} {\bibinfo  {journal} {Phys. Rev. X}\ }\textbf {\bibinfo {volume}
  {5}},\ \bibinfo {pages} {041002} (\bibinfo {year} {2015})}\BibitemShut
  {NoStop}%
\bibitem [{\citenamefont {Yuan}\ \emph
  {et~al.}(2015{\natexlab{a}})\citenamefont {Yuan}, \citenamefont {Singh},
  \citenamefont {Blanter},\ and\ \citenamefont {Steele}}]{ref:supp_yuan2015lcm}%
  \BibitemOpen
  \bibfield  {author} {\bibinfo {author} {\bibfnamefont {M.}~\bibnamefont
  {Yuan}}, \bibinfo {author} {\bibfnamefont {V.}~\bibnamefont {Singh}},
  \bibinfo {author} {\bibfnamefont {Y.~M.}\ \bibnamefont {Blanter}}, \ and\
  \bibinfo {author} {\bibfnamefont {G.~A.}\ \bibnamefont {Steele}},\
  }\href@noop {} {\bibfield  {journal} {\bibinfo  {journal} {Nat. Commun.}\
  }\textbf {\bibinfo {volume} {6}} (\bibinfo {year}
  {2015}{\natexlab{a}})}\BibitemShut {NoStop}%
\bibitem [{\citenamefont {Purdy}\ \emph {et~al.}(2012)\citenamefont {Purdy},
  \citenamefont {Peterson}, \citenamefont {Yu},\ and\ \citenamefont
  {Regal}}]{ref:supp_purdy2012cos}%
  \BibitemOpen
  \bibfield  {author} {\bibinfo {author} {\bibfnamefont {T.~P.}\ \bibnamefont
  {Purdy}}, \bibinfo {author} {\bibfnamefont {R.~W.}\ \bibnamefont {Peterson}},
  \bibinfo {author} {\bibfnamefont {P.-L.}\ \bibnamefont {Yu}}, \ and\ \bibinfo
  {author} {\bibfnamefont {C.~A.}\ \bibnamefont {Regal}},\ }\href@noop {}
  {\bibfield  {journal} {\bibinfo  {journal} {New J. Phys.}\ }\textbf {\bibinfo
  {volume} {14}},\ \bibinfo {pages} {115021} (\bibinfo {year}
  {2012})}\BibitemShut {NoStop}%
\bibitem [{\citenamefont {Yuan}\ \emph
  {et~al.}(2015{\natexlab{b}})\citenamefont {Yuan}, \citenamefont {Cohen},\
  and\ \citenamefont {Steele}}]{ref:supp_yuan2015snm}%
  \BibitemOpen
  \bibfield  {author} {\bibinfo {author} {\bibfnamefont {M.}~\bibnamefont
  {Yuan}}, \bibinfo {author} {\bibfnamefont {M.~A.}\ \bibnamefont {Cohen}}, \
  and\ \bibinfo {author} {\bibfnamefont {G.~A.}\ \bibnamefont {Steele}},\
  }\href@noop {} {\bibfield  {journal} {\bibinfo  {journal} {Appl. Phys.
  Lett.}\ }\textbf {\bibinfo {volume} {107}},\ \bibinfo {eid} {263501}
  (\bibinfo {year} {2015}{\natexlab{b}})}\BibitemShut {NoStop}%
\end{thebibliography}
